\documentclass[preprint,nofootinbib,aps,superscriptaddress,eqsecnum]{revtex4-2}
\usepackage{float}
\usepackage{extarrows}
\usepackage{tabularx}
\usepackage{array}
\usepackage{graphicx}
\usepackage{amsmath}
\usepackage{multirow}
\usepackage{amsfonts}
\usepackage{amssymb}
\usepackage{hyperref}
\usepackage[justification=centering]{caption} 
\usepackage[margin=1in]{geometry}
\usepackage{graphicx,subcaption,lipsum}
\let\revappendix\appendix
\usepackage{caption}
\usepackage{subcaption}
\usepackage{color}
\allowdisplaybreaks
\def\bea{\begin{eqnarray}}
\def\eea{\end{eqnarray}}
\def\be{\begin{equation}}
\def\ee{\end{equation}}

\begin{document}
\title{Neutrino Textures from Modular $A_4$ Left--Right Symmetry:
Experimental Signatures at DUNE and T2HK in the Post-JUNO Era}
\author{Bhabana Kumar}
\email{bhabanakumar474@gmail.com}
\affiliation{Department of Physics, Tezpur University, Assam 784028, India}

\author{Debajyoti Dutta}
\email{phy.debajyoti@bhattadevuniversity.ac.in}
\affiliation{Department of Physics, Bhattadev University, Assam 781325, India}

\author{Mrinal Kumar Das}
\email{mkdas@tezu.ernet.in}
\affiliation{Department of Physics, Tezpur University, Assam 784028, India}

\begin{abstract}
We have realized different two-zero textures within the framework of the left right symmetric model using the $\Gamma_{3}\cong A_{4}$ modular group. The matter multiplets of the model are assigned as three singlet representations of the $A_{4}$ group, and their charge assignments together with the modular weights of the Yukawa couplings are chosen in such a way that different two-zero textures of the neutrino mass matrix are obtained. In total, we have successfully realized seven different two-zero textures. Furthermore, we have studied neutrinoless double beta decay and lepton flavor violating (LFV) processes, and have calculated the effective Majorana mass and the branching ratios for  LFV processes for each of the textures. We further probe these two-zero textures at the long-baseline neutrino experiments DUNE and T2HK. We find that DUNE, especially when combined with T2HK, can significantly restrict the $\theta_{23}$–$\delta_{\rm CP}$ parameter space predicted by these textures. Moreover, the inclusion of high-precision determinations of $\theta_{12}$ (from JUNO) and $\theta_{13}$ leads to a substantial, further reduction of the allowed parameter space. For assumed inverted mass ordering, the synergy of DUNE and T2HK,  leads to a highly predictive scenario for the $B_{2}$ and $B_{4}$ textures, as the allowed regions collapse into tiny islands near the CP-conserving points in the lower and higher octant of $\theta_{23}$, respectively. 
\end{abstract}
\maketitle
\newpage

\section{\textbf{Introduction}}
Fermion flavor pattern is one of the outstanding problems in particle physics. So far, the quark mass and mixing parameters have been measured precisely, but in the neutral lepton sector, many questions still remain unanswered. From neutrino oscillation experiments, we know that neutrinos have small and non-degenerate masses, and lepton mixing is possible. Both quark and lepton mixings are possible, but we observe a significant hierarchy problem. Neutrino masses are of the order of $\mathcal{O}(\text{eV})$, whereas the top quark mass lies at the TeV scale. This large mass difference between leptons and quarks remains an open question in flavor physics. Furthermore, the CKM mixing matrix in the quark sector is almost diagonal, exhibiting small mixing angles. In contrast, the PMNS mixing matrix in the lepton sector contains two large mixing angles known as atmospheric and solar mixing angles and one small mixing angle, referred to as the reactor mixing angle. The PMNS matrix is therefore significantly different from the CKM matrix.\\
The oscillation of the three active neutrinos can be described by six parameters: two mass-squared differences, three mixing angles, and one Dirac CP phase. Different neutrino oscillation experiments have precisely measured the mass-squared differences ($\Delta m^2_{\text{sol}}$, $\Delta m^2_{\text{atm}}$) and the three mixing angles $(\theta_{12}, \theta_{13}, \theta_{23})$, but the Dirac CP phase has not been determined precisely so far. Additionally, we still do not know the absolute neutrino mass scale, the nature of the mass hierarchy, or whether neutrinos are Majorana or Dirac particles. If neutrinos are Majorana particles, then the neutrino mass matrix is a $3 \times 3$ symmetric matrix, which contains 12 real parameters. Among these 12 free parameters, three phases can be absorbed into the redefinition of the charged lepton fields. As a result, we are left with nine physical parameters, but only six of them can be constrained by current neutrino oscillation data. Thus, the flavor problem in particle physics remains unresolved, and various theoretical approaches have been proposed to address it. One such approach is the texture-zero hypothesis, where the number of free parameters in the fermion mass matrix is reduced by assuming that some elements of the mass matrix are zero. The concept of texture zeros in both quark and lepton mass matrices has been extensively studied in the literature\cite{Low:2004wx,Low:2005yc,Ludl:2014axa,Grimus:2004hf,Dev:2011jc}. It has been shown that a neutrino mass matrix with more than two independent zeros is not consistent with the current oscillation data\cite{Xing:2004ik}. However, a neutrino mass matrix with two independent zeros commonly referred to as a two-zero texture has only five free parameters. Since the number of free parameters is fewer than the number of available experimental observables, two-zero textures are particularly attractive. There are 15 possible two-zero textures for the neutrino mass matrix, but only seven of them are compatible with current neutrino oscillation data.\\
Since the predictivity of pure two-zero texture is weak~\cite{Xing:2003ic, Kitabayashi:2019uzg,Merle:2006du}, one of the straightforward ways to impose a vanishing element into the neutrino mass matrix can be obtained by using a non-Abelian discrete symmetric group\cite{Grimus:2004az,Borgohain:2018lro,Dev:2011jc}. However, recently the use of modular symmetry in model building has gained popularity over discrete flavor symmetries due to the advantage that in modular symmetry, the flavor symmetry can be broken without introducing any flavon or extra field. This is because, in the case of modular symmetry, the modulus $\tau$ is responsible for breaking the flavor symmetry. In models constructed using modular symmetry, the Yukawa couplings correspond to modular forms of level $N$, and these Yukawa couplings are holomorphic functions of the complex modulus field $\tau$. Due to the advantage of the modular group over discrete symmetries, nowadays the modular group is widely used in model building, including the construction of neutrino mass models based on $\Gamma_{3}$ \cite{Nomura:2019xsb,Gogoi:2022jwf,Kumar:2023moh,Kashav:2022kpk,Kashav:2021zir,Behera:2020lpd,deAnda:2018ecu},$\Gamma_{4}$ \cite{Okada:2019lzv,Novichkov:2020eep,Penedo:2018nmg,Ding:2021zbg}, and $\Gamma_{5}$ \cite{Novichkov:2018nkm}, and in the study of texture-zero structures~\cite{Zhang:2019ngf,Kikuchi:2022svo,Ding:2022aoe,Nomura:2023usj,Nomura:2024ghc,Treesukrat:2025dhd}.
In this work, we have employed the $\Gamma_{3}$ modular group to realize different two-zero textures and the light neutrino masses are generated via an extended type II seesaw mechanism. After constructing the various two-zero textures, we have calculated the effective Majorana mass and branching ratios for three lepton flavor violating processes. In addition, a key objective of this work is to probe the viability of allowed two-zero neutrino mass textures at long-baseline experiments such as DUNE\cite{DUNE:2020lwj,DUNE:2020ypp,DUNE:2020txw,DUNE:2022aul} and T2HK\cite{Hyper-Kamiokande:2016srs}. Earlier, phenomenology of the Majorana neutrino textures have been studied in the context of DUNE at \cite{Bora:2016ygl, Borah:2025vtn}. Here in this work, we have performed a detailed numerical analysis using GLoBES \cite{Huber:2004ka,Huber:2007ji} to fit the oscillation parameters predicted by these textures and to evaluate their sensitivities at DUNE, as well as in the combined DUNE+T2HK configuration. Our results show that, within the current framework of neutrino oscillation physics, DUNE alone can exclude sizable regions of the $\theta_{23}$–$\delta_{\rm CP}$ parameter space implied by these textures. The inclusion of T2HK in combination with DUNE further improves the sensitivity, leading to significantly stronger constraints on this parameter space. In addition, recent data from the JUNO experiment \cite{JUNO:2025gmd}, corresponding to 60 days of operation, have led to a substantial improvement in the precision of the solar mixing angle $\theta_{12}$. The impact of a precise determination of $\theta_{12}$ has been investigated in Refs.~\cite{Borah:2025vtn, Araya-Santander:2025jfd, Ding:2025dqd, Goswami:2025wla, Capozzi:2025ovi, Ge:2025cky, Petcov:2025aci, Bora:2025xfj, Ding:2025dqd} within the framework of various extensions of neutrino physics. We have also studied the impact of this precision measurement on the capability of long-baseline experiments to probe two-zero textures predicted by the
modular $A_4$ left–right symmetric model.\\
The paper is organized in the following way: In Section II, we briefly describe the left-right symmetric model with extended type II seesaw. In Sections III, we provide brief introductions to the two-zero texture of the neutrino mass matrix. Section IV contains a detailed description of the model, including the charge assignments and modular weights assigned to the supermultiplets to realize different two-zero textures. A brief review of DUNE and T2HK is given in Section V. Finally, the numerical analysis and conclusions are presented in Sections VI and VII, respectively.
\section{\textbf{Left right symmetric model with Extended Type II seesaw mechanism}}\label{s1}
The left-right symmetric model is based on the gauge group $SU(2)_{L} \times SU(2)_{R} \times U(1)_{B-L} \times SU(3)_{C}$, denoted as $\mathcal{G}_{2213}$\cite{Mohapatra:1974gc,Senjanovic:1975rk,Mohapatra:1977mj}. The particle content and their charge assignments under the gauge group $\mathcal{G}_{2213}$ in the minimal SUSY LRSM are given below \cite{Aulakh:1998nn,Borah:2010kk,Babu:2008ep,Aulakh:1997ba}
\begin{align*}
    L_{L} &= \begin{pmatrix}
        \nu_{L} \\
        e_{L}
    \end{pmatrix} \sim (2,1,-1,1), \hspace{1cm}
    L^{c}_{R} = \begin{pmatrix}
        \nu_{R}^{c}\\
        e_{R}^{c}
    \end{pmatrix} \sim (1,2,1,1) \\[10pt]
    q &= \begin{pmatrix}
        u_{L}\\
        d_{L}
    \end{pmatrix} \sim (2,1,\tfrac{1}{3},3), \hspace{1cm}
    q^{c}_{R} = \begin{pmatrix}
        u_{R}^{c} \\
        d_{R}^{c}
    \end{pmatrix} \sim (1,2,-\frac{1}{3},3)
\end{align*}
The scalar sector of the minimal SUSY LRSM is given by
\begin{align*}
    \Phi_{1} &= \begin{pmatrix}
         \phi^{0}_{11} & \phi^{+}_{11} \\
         \phi^{-}_{12} & \phi^{0}_{12}
    \end{pmatrix} 
    &&\sim (2,2,0,1), 
    &\hspace{1cm}
    \Phi_{2} &= \begin{pmatrix}
         \phi^{0}_{21} & \phi^{+}_{21} \\
         \phi^{-}_{22} & \phi^{0}_{22}
    \end{pmatrix}
    &&\sim (2,2,0,1) \\[10pt]
    \Delta_{L} &= \begin{pmatrix}
         \tfrac{1}{2}\delta^{+}_{L} & \delta^{++}_{L} \\
         \delta^{0}_{L} & \tfrac{1}{2}\delta^{+}_{L}
    \end{pmatrix}
    &&\sim (3,1,2,1), 
    &\hspace{1cm}
    \overline{\Delta_{L}} &= \begin{pmatrix}
         \tfrac{1}{2}\delta^{-}_{L} & \delta^{0}_{L} \\
         \delta^{--}_{L} & -\tfrac{1}{2}\delta^{-}_{L}
    \end{pmatrix}
    &&\sim (3,1,-2,1) \\[10pt]
     \Delta_{R}^{c} &= \begin{pmatrix}
         \tfrac{1}{2}\delta^{-}_{R} & \delta^{--}_{R} \\
         \delta^{0}_{R} & -\tfrac{1}{2}\delta^{-}_{R}
    \end{pmatrix}
    &&\sim (1,3,-2,1), 
    &\hspace{1cm}
    \overline{\Delta_{R}^{c}} &= \begin{pmatrix}
         \tfrac{1}{2}\delta^{+}_{R} & \delta^{0}_{R} \\
         \delta^{++}_{R} & -\tfrac{1}{2}\delta^{+}_{R}
    \end{pmatrix}
    &&\sim (1,3,2,1)
\end{align*}
In this model, neutrino masses arise through an extended Type-II seesaw mechanism by introducing one sterile fermion $S_i$ per generation. The scalar sector consists of a bidoublet, a scalar triplet, and a scalar doublet. The complete particle content and their charge assignments under the gauge group of LRSM are summarized in Table. 1.
\begin{table}[H]
\centering
\begin{tabular}{|p{2cm}|c|c|c|c|c|c|c|c|}
\hline

Field & $L^{c}_{R_{i}}$ & $L_{L_{i}}$ &$S$ & $\Phi_{i}$ & $\Delta^{c}_{R}$ & $\chi^{c}_{R}$ & $\Delta_{L}$&$\chi_{L}$ \\ \hline

$SU(2)_L$ & 1 & 2 &  1 & 2 & 1 & 1 & 3& 2 \\ \hline
$SU(2)_R$ & 2 & 1  & 1 &  2 & 3 & 2 & 1 & 1 \\ \hline
$U(1)_{B-L}$ & 1 & -1  & 0 & 0 & -2 & -1 & 2 & 1 \\ \hline

\end{tabular}
\caption{Charge assignment under $SU(2)_{L} \times SU(2)_{R} \times U(1)_{B-L}$ for the particle content of the model}
\end{table}
The superpotential associated with the model is given in equation \eqref{W3Q1}
\begin{equation}\label{W3Q1}
\begin{aligned}
\mathcal{W} &= 
Y_{L} \, L^{T}_{L} \, i\sigma_{2} \, \Phi_{i} \, L^{c}_{R}
+ f_{L} \, L^{T}_{L} \, i\sigma_{2} \, \Delta_{L} \, L_{L} 
+ f_{R} \, L^{c\,T}_{R} \, i\sigma_{2} \, \Delta^{c}_{R} \, L^{c}_{R} \\
&\quad 
+ Y_{SL} \, L^{T}_{L} \, i\sigma_{2} \, \chi_{L} \, S
+ Y_{SR} \, L^{c\,T}_{R} \, i\sigma_{2} \, \chi^{c}_{R} \, S
+ \mu_{S} \, S \, S \\[3pt]
&\supset 
M_{D} \, \nu^{T}_{L} \, \nu^{c}_{R}
+ M_{L} \, \nu^{T}_{L} \, \nu_{L}
+ M_{R} \, \nu^{c\,T}_{R} \, \nu^{c}_{R} \\
&\quad 
+ M \, \nu^{c\,T}_{R} \, S
+ \mu_{L} \, \nu^{c\,T}_{L} \, S
+ \mu_{S} \, S \, S
\end{aligned}
\end{equation}
The term $\mu_s\, SS$ is, in principle, arbitrary; however, we set it to zero in this analysis. We also take $\langle \chi_L \rangle \to 0$. After spontaneous symmetry breaking of the superpotential in Eq.~\eqref{W3Q1}, the resulting $9\times 9$ mass matrix can be written as follows.
\begin{equation}\label{W3Q2}
    \mathcal{M}= \begin{pmatrix}
                   M_{L} & 0 & M_{D} \\
                   0 & 0 & M \\
                   M^{T}_{D} & M^{T} & M_{R}
                \end{pmatrix}
\end{equation}
Assuming the mass hierarchy $M_R > M > M_D \gg M_L$, and performing block diagonalization of the matrix $\mathcal{M}$, the resulting mass matrices for the active, right-handed, and sterile neutrinos are obtained as follows.
\begin{align}\label{W2Q3}
    m_{\nu} &= M_L, \nonumber \\
    M_N &= M_R = \frac{v_R}{v_L} M_L, \nonumber \\
    M_S &= -M M_R^{-1} M^T ~~.
\end{align}
The complete $9 \times 9$ unitary matrix responsible for diagonalizing the matrix $\mathcal{M}$, given in equation~\ref{W3Q2}, can be written in the following way.
\begin{equation}\label{W3Q4}
\mathcal{V} = 
\left(
\begin{array}{ccc}
V^{\nu\nu} & V^{\nu S} & V^{\nu N} \\
V^{S\nu} & V^{SS} & V^{SN} \\
V^{N\nu} & V^{NS} & V^{NN}
\end{array}
\right)
=
\left(
\begin{array}{ccc}
U_{\nu} & M_{D} M^{-1} U_{S} & M_{D} M_{R}^{-1} U_{N} \\
(M_{D} M^{-1})^{\dagger} U_{\nu} & U_{S} & M M_{R}^{-1} U_{N} \\
0 & -(M M_{R}^{-1})^{\dagger} U_{S} & U_{N}
\end{array}
\right)
\end{equation}
where $U_{\nu}$, $U_{S}$ and $U_{N}$ are $3\times 3$ matrix responsible for diagonalizing the active neutrino, sterile neutrino and right handed neutrino mass matrix, respectively. The matrix $U_{\nu}$ corresponds to the $U_{PMNS}$ matrix. This $U_{PMNS}$ matrix can be parametrised by using three mixing angles and three phases. The structure of this $U_{PMNS}$ matrix is given in equation \ref{W3q5}
\begin{equation}\label{W3q5}
U_{PMNS} =
\begin{pmatrix}
    c_{12}c_{13} & s_{12}c_{13} & s_{13}e^{-i\delta_{CP}} \\
    -s_{12}c_{23} - c_{12}s_{23}s_{13}e^{i\delta_{CP}} &
    c_{12}c_{23} - s_{12}s_{23}s_{13}e^{i\delta_{CP}} &
    s_{23}c_{13} \\
    s_{12}s_{23} - c_{12}c_{23}s_{13}e^{i\delta_{CP}} &
    -c_{12}s_{23} - s_{12}c_{23}s_{13}e^{i\delta_{CP}} &
    c_{23}c_{13}
\end{pmatrix}
\begin{pmatrix}
    1 & 0 & 0 \\
    0 & e^{i\alpha} & 0 \\
    0 & 0 & e^{i\beta}
\end{pmatrix}~.
\end{equation}
where $c_{ij}$ and $s_{ij}$ represent $\cos\theta_{ij}$ and $\sin\theta_{ij}$, respectively. $\delta_{CP}$ is the Dirac CP-violating phase, and $\alpha$ and $\beta$ are the two Majorana phases.  The $3\sigma$ ranges of these neutrino oscillation parameters are given in \cite{Esteban:2024eli}
\section{\textbf{Two zero Texture of neutrino mass matrix}}\label{s2}
Considering that the charged lepton mass matrix is diagonal and the neutrino is a Majorana particle, it is possible to have fifteen two-zero textures in the neutrino mass matrix. Among these, only seven textures are experimentally compatible. These seven two-zero textures can be classified into three distinct classes: class A, which includes $A_{1}$ and $A_{2}$; class B, which includes $B_{1}$, $B_{2}$, $B_{3}$, and $B_{4}$; and finally, class C. The structures of the experimentally viable two-zero textures of the Majorana mass matrix are listed below.
\[
\begin{array}{cc}
\textbf{Class } A_{1} &
\hspace{2.5cm} \textbf{Class } A_{2} \\
\begin{pmatrix}
0 & 0 & \times \\
0 & \times & \times \\
\times & \times & \times
\end{pmatrix}
& 
\hspace{2.5cm}
\begin{pmatrix}
0 & \times & 0 \\
\times & \times & \times \\
0 & \times & \times
\end{pmatrix}
\end{array}
\]
\[
\begin{array}{cccc}
\textbf{Class } B_{1} &
\hspace{1.5cm} \textbf{Class } B_{2} & \hspace{1.5cm} \textbf{Class } B_{3} &\hspace{1.5cm} \textbf{Class } B_{4} \\
\begin{pmatrix}
\times & \times & 0 \\
\times & 0 & \times \\
0 & \times & \times
\end{pmatrix}
& 
\hspace{1.5cm}
\begin{pmatrix}
\times & 0 & \times \\
0 & \times & \times \\
\times & \times & 0
\end{pmatrix}
& 
\hspace{1.5cm}
\begin{pmatrix}
\times & 0 & \times \\
0 & 0 & \times \\
\times & \times & \times
\end{pmatrix}
& 
\hspace{1.5cm}
\begin{pmatrix}
\times & \times & 0 \\
\times & \times & \times \\
0 & \times & 0
\end{pmatrix}
\end{array}
\]
\[
\begin{array}{c}
\textbf{Class } C\\
\begin{pmatrix}
\times & \times & \times \\
\times & 0 & \times \\
\times & \times & 0
\end{pmatrix}
\end{array}
\]
in which "$\times$" represents a non-zero entry in the corresponding position. 
\section{\textbf{The Model}}\label{s3}
Initially, we have considered the lepton doublets to transform as $1$, $1^{\prime}$, and $1^{\prime\prime}$, and the right-handed charged leptons to transform as $1$, $1^{\prime\prime}$, and $1^{\prime}$. The modular weights are chosen in such a way that we obtain a diagonal charged lepton mass matrix from the interaction term $Y_{l}(L^{T}_{L}i\sigma_{2} \Phi L^{c}_{R})$. At the same time, we obtain a two-zero texture of the neutrino mass matrix from the interaction term $F_{L}(L^{T}_{L} i\sigma_{2}\Delta_{L} L_{L})$. To generate the other two-zero textures, we have interchanged the charge assignments as well as the modular weights among the left-handed lepton doublets, such that different two-zero textures are obtained. Simultaneously, the charge assignments and modular weights of $L^{c}_{R}$ have also been modified appropriately to ensure that the charged lepton mass matrix remains diagonal in each case.  All the scalar fields are considered to transform as trivial singlets under the $A_{4}$ symmetry, and their modular weights are taken to be zero. The sterile fermions $S_{i}$ transform as a triplet under the $A_{4}$ group.
\subsection{\textbf{Class C}}
To obtain the texture C, we have considered the left-handed lepton doublets $L_{L_{i}}$ to transform as $1$, $1^{\prime}$, and $1^{\prime\prime}$, and assigned them modular weights $-5$, $-3$, and $-3$, respectively. At the same time, in order to obtain a diagonal charged lepton mass matrix, we have taken the right-handed charged leptons $L^{c}_{R_{i}}$ to transform as $1$, $1^{\prime\prime}$, and $1^{\prime}$, carrying modular weights $-3$, $-1$, and $3$, respectively. The charge assignments and modular weights for all the particle contents of the model are summarized in Table~\ref{W3T3}.
\begin{table}[ht]
\centering
\begin{tabular}{|p{2cm}|c|c|c|c|c|c|c|c|c|}
\hline
Field & $L^{c}_{R_{i}}$ & $L_{L_{i}}$ & $S$ & $\Phi$ & $\Delta^{c}_{R}$ & $\chi^{c}_{R}$ & $\Delta_{L}$&$\chi_{L}$ \\ \hline
$A_{4}$ & $1,1^{\prime\prime},1^{\prime}$ & $1,1^{\prime},1^{\prime\prime}$ & $3$ & $1$ & $1$ & $1$ & $1$ & $1$\\ \hline
$K_{I}$ & $-3,-1,3$ & $-5,-3,-3$ & 1 & 0 & 0 & 0 & 0 & 0 \\ \hline
\end{tabular}
\caption{Charge assignment and modular weights under $A_{4}$ for the particle content of the model}
\label{W3T3}
\end{table}
 The superpotential is given in equations~\eqref{W3Q5} and~\eqref{W3Q7}, and the Majorana mass matrices for the light neutrinos and the right-handed neutrinos are presented in equations~\eqref{W3Q6} and~\eqref{W3Q8}, respectively.
\begin{equation}\label{W3Q5}
\begin{aligned}
\mathcal{W}_{L} =\ & 
a_{11}\Delta_{L} L^{T}_{L_{1}} Y^{10}_{1} L_{L_{1}} 
+ a_{12}(\Delta_{L} L^{T}_{L_{1}} Y^{8}_{1^{\prime\prime}} L_{L_{2}}+\Delta_{L} L^{T}_{L_{2}} Y^{8}_{1^{\prime\prime}} L_{L_{1}} ) \\
&+a_{13} (\Delta_{L} L^{T}_{L_{1}} Y^{8}_{1^{\prime}} L_{L_{3}}+ \Delta_{L} L^{T}_{L_{3}} Y^{8}_{1^{\prime}} L_{L_{1}}  )
+a_{23}( \Delta_{L} L^{T}_{L_{2}} Y^{6}_{1} L_{L_{3}} 
 + \Delta_{L} L^{T}_{L_{3}} Y^{6}_{1} L_{L_{2}})
\end{aligned}
\end{equation}
\begin{equation}\label{W3Q6}
    M_{L}=v_{L}\begin{pmatrix}
             a_{11} Y^{10}_{1} & a_{12} Y^{8}_{\prime\prime} & a_{13}Y^{8}_{1^{\prime}}\\
             a_{12} Y^{8}_{\prime\prime} & 0 & a_{23} Y^{6}_{1} \\
             a_{13}Y^{8}_{1^{\prime}} & a_{23}Y^{6}_{1} & 0
           \end{pmatrix}
\end{equation}
Here, $v_{L}$ is the VEV of the scalar triplet $\Delta_{L}$, and its value typically lies in the sub-eV range. The parameters $a_{11}$, $a_{12}$, $a_{13}$, and $a_{23}$ are free parameters. $M_{L}$ denotes the mass matrix for active neutrinos, and the texture obtained for it corresponds to the two-zero texture of class C.
\begin{equation}\label{W3Q7}
\begin{aligned}
\mathcal{W}_{R} =\ & 
a_{11}\Delta^{c}_{R} L^{c^T}_{R_{1}} Y^{10}_{1} L^{c}_{R_{1}} 
+ a_{12}(\Delta^{c}_{R} L^{c^T}_{R_{1}} Y^{8}_{1^{\prime\prime}} L^{c}_{R_{2}}+\Delta^{c}_{R} L^{c^T}_{R_{2}} Y^{8}_{1^{\prime\prime}} L^{c}_{R_{1}} ) \\
&+a_{13} (\Delta^{c}_{R} L^{c^T}_{R_{1}} Y^{8}_{1^{\prime}} L^{c}_{R_{3}}+ \Delta^{c}_{R} L^{c^T}_{R_{3}} Y^{8}_{1^{\prime}} L^{c}_{R_{1}}  )
+a_{23}( \Delta^{c}_{R} L^{c^T}_{R_{2}} Y^{6}_{1} L^{c}_{R_{3}} 
 + \Delta^{c}_{R} L^{c^T}_{R_{3}} Y^{6}_{1} L^{c}_{R_{2}})
\end{aligned}
\end{equation}

\begin{equation}\label{W3Q8}
      M_{R}=\frac{v_{R}}{v_{L}}\begin{pmatrix}
             a_{11} Y^{10}_{1} & a_{12} Y^{8}_{\prime\prime} & a_{13}Y^{8}_{1^{\prime}}\\
             a_{12} Y^{8}_{\prime\prime} & 0 & a_{23} Y^{6}_{1} \\
             a_{13}Y^{8}_{1^{\prime}} & a_{23}Y^{6}_{1} & 0
           \end{pmatrix}
\end{equation}
The $A_{4}$ invariant superpotential for the neutrino sterile mixing is given in equation ~\eqref{W3Q9}, and its corresponding mass matrix is given in the equation ~\eqref{W3Q10}
\begin{equation}\label{W3Q9}
    \mathcal{W}_{NS}= g_{1}\chi^{c}_{R} L^{c^T}_{R_{1}}(Y^{4}_{3}S)_{1} + g_{2}\chi^{c}_{R} L^{c^T}_{R_{2}}(Y^{2}_{3}S)_{1^{\prime\prime}}+ g_{3}\chi^{c}_{R} L^{c^T}_{R_{3}}(Y^{2}_{3}S)_{1^{\prime}}
\end{equation}
\begin{equation}\label{W3Q10}
    M = v^{\prime}\begin{pmatrix}
               g_{1}(Y^{2}_{1}-Y_{2}Y_{3}) & g_{1}(Y^{2}_{2}-Y_{1}Y_{3}) & g_{1}(Y^{2}_{3}-Y_{1}Y_{2}) \\
               g_{2}Y_{3} & g_{2}Y_{2} & g_{2}Y_{1}\\
               g_{3}Y_{2} & g_{3}Y_{1} & g_{3}Y_{3}
              \end{pmatrix}
\end{equation}
\subsection{\textbf{Class $B_{1}$, $B_{2}$, $B_{3}$, and $B_{4}$}}
\begin{table}[ht]
\centering
\resizebox{\textwidth}{!}{
\begin{tabular}{|p{2.2cm}|c|c|c|c|c|c|c|c|}
\hline
\textbf{Field} & $L^{c}_{R_{i}}$ & $L_{L_{i}}$ & $S$ & $\Phi$ & $\Delta^{c}_{R}$ & $\chi^{c}_{R}$ & $\Delta_{L}$ & $\chi_{L}$ \\ \hline

\multicolumn{9}{|c|}{\textbf{Texture $B_1$}} \\ \hline
$A_{4}$ & $1,1^{\prime},1^{\prime\prime}$ & $1,1^{\prime\prime},1^{\prime}$ & $3$ & $1$ & $1$ & $1$ & $1$ & $1$ \\
$k_{I}$ & $2,2,2$ & $-2,-2,-2$ & 0 & 0 & 0 & 0 & 0 & 0 \\ \hline

\multicolumn{9}{|c|}{\textbf{Texture $B_2$}} \\ \hline
$A_{4}$ & $1,1^{\prime\prime},1^{\prime}$ & $1,1^{\prime},1^{\prime\prime}$ & $3$ & $1$ & $1$ & $1$ & $1$ & $1$ \\
$k_{I}$ & $2,2,2$ & $-2,-2,-2$ & 0 & 0 & 0 & 0 & 0 & 0 \\ \hline

\multicolumn{9}{|c|}{\textbf{Texture $B_3$}} \\ \hline
$A_{4}$ & $1,1^{\prime},1^{\prime\prime}$ & $1,1^{\prime\prime},1^{\prime}$  & $3$ & $1$ & $1$ & $1$ & $1$ & $1$ \\
$k_{I}$ & $0,-2,-2$ & $-4,-2,-4$ & 0 & 0 & 0 & 0 & 0 & 0 \\ \hline

\multicolumn{9}{|c|}{\textbf{Texture $B_4$}} \\ \hline
$A_{4}$ & $1,1^{\prime\prime},1^{\prime}$ & $1,1^{\prime},1^{\prime\prime}$ & $3$ & $1$ & $1$ & $1$ & $1$ & $1$ \\
$k_{I}$ & $0,-2,-2$ & $-4,-4,-2$  & 0 & 0 & 0 & 0 & 0 & 0 \\ \hline
\end{tabular}}
\caption{Charge assignments and modular weights under $A_{4}$ for the particle content of the model for textures $B_1$, $B_2$, $B_3$, and $B_4$.}
\label{W3T4}
\end{table}
To obtain the $B_{2}$ texture of the active neutrino mass matrix, the same charge assignments are given to all the particle contents of the model, but different modular weights are assigned to those particles. The charge assignments and modular weights considered to generate the $B_{2}$ texture of the neutrino mass matrix are presented in Table~\ref{W3T4}. To achieve the $B_{1}$ texture of the neutrino mass matrix, we have interchanged the charge assignments and modular weights of $L_{L_{2}}$ and $L_{L_{3}}$. Additionally, to maintain a diagonal charged lepton mass matrix, we have also interchanged the charge assignments and modular weights of $L^{c}_{R_{2}}$ and $L^{c}_{R_{3}}$.
The $A_{4}$-invariant superpotentials for the left-handed charged leptons corresponding to textures $B_{1}$, $B_{2}$, $B_{3}$, and $B_{4}$ are given in equations~\eqref{W3Q11a}, \eqref{W3Q11b}, \eqref{W3Q11c}, and \eqref{W3Q11d}, respectively.
\begin{subequations}\label{W3Q11}
\begin{align}
  \mathcal{W}^{(B_{1})}_{L} =\ & a_{11} \Delta_{L} L^{T}_{L_{1}} Y^{4}_{1} L_{L_{1}} 
  + a_{12} (\Delta_{L} L^{T}_{L_{1}} Y^{4}_{1^{\prime}} L_{L_{2}} + \Delta_{L} L^{T}_{L_{2}} Y^{4}_{1^{\prime}} L_{L_{1}}) \notag\\
  & + a_{23} (\Delta_{L} L^{T}_{L_{2}} Y^{4}_{1} L_{L_{3}} + \Delta_{L} L^{T}_{L_{3}} Y^{4}_{1} L_{L_{2}}) 
  + a_{33} \Delta_{L} L^{T}_{L_{3}} Y^{4}_{1^{\prime}} L_{L_{3}} \label{W3Q11a} \\[1ex]
  \mathcal{W}^{(B_{2})}_{L} =\ & a_{11} \Delta_{L} L^{T}_{L_{1}} Y^{4}_{1} L_{L_{1}} 
  + a_{13} (\Delta_{L} L^{T}_{L_{1}} Y^{4}_{1^{\prime}} L_{L_{3}} + \Delta_{L} L^{T}_{L_{3}} Y^{4}_{1^{\prime}} L_{L_{1}}) \notag\\
  & + a_{22} \Delta_{L} L^{T}_{L_{2}} Y^{4}_{1} L_{L_{2}} 
  + a_{23} (\Delta_{L} L^{T}_{L_{2}} Y^{4}_{1} L_{L_{3}} + \Delta_{L} L^{T}_{L_{3}} Y^{4}_{1} L_{L_{2}}) \label{W3Q11b} \\[1ex]
  \mathcal{W}^{(B_{3})}_{L} =\ & a_{11} \Delta_{L} L^{T}_{L_{1}} Y^{8}_{1} L_{L_{1}} 
  + a_{13} (\Delta_{L} L^{T}_{L_{1}} Y^{8}_{1^{\prime\prime}} L_{L_{3}} + \Delta_{L} L^{T}_{L_{3}} Y^{8}_{1^{\prime\prime}} L_{L_{1}}) \notag\\
  & + a_{23} (\Delta_{L} L^{T}_{L_{2}} Y^{6}_{1} L_{L_{3}} + \Delta_{L} L^{T}_{L_{3}} Y^{6}_{1} L_{L_{2}}) 
  + a_{33} \Delta_{L} L^{T}_{L_{2}} Y^{8}_{1^{\prime}} L_{L_{2}} \label{W3Q11c} \\[1ex]
  \mathcal{W}^{(B_{4})}_{L} =\ & a_{11} \Delta_{L} L^{T}_{L_{1}} Y^{8}_{1} L_{L_{1}} 
  + a_{12} (\Delta_{L} L^{T}_{L_{1}} Y^{8}_{1^{\prime\prime}} L_{L_{2}} + \Delta_{L} L^{T}_{L_{2}} Y^{4}_{1^{\prime\prime}} L_{L_{1}}) \notag\\
  & + a_{22} \Delta_{L} L^{T}_{L_{2}} Y^{8}_{1^{\prime}} L_{L_{2}} 
  + a_{23} (\Delta_{L} L^{T}_{L_{2}} Y^{6}_{1} L_{L_{3}} + \Delta_{L} L^{T}_{L_{3}} Y^{6}_{1} L_{L_{2}}) \label{W3Q11d}
\end{align}
\end{subequations}
The four textures of class B, obtained from the superpotential associated with the lepton–lepton interaction, are given in Table~\ref{W3T5}.
\begin{table}[htbp]
\centering
\renewcommand{\arraystretch}{1.7}
\resizebox{\textwidth}{!}{
\begin{tabular}{|c|c|c|c|c|}
\hline
\textbf{Matrix} & \textbf{\( B_1 \)} & \textbf{\( B_2 \)} & \textbf{\( B_3 \)} & \textbf{\( B_4 \)} \\
\hline
\( M_L \) &
\( v_L\begin{pmatrix}
a_{11}Y_1^4 & a_{12}Y_{1'}^4 & 0\\
a_{12}Y_{1'}^4 & 0 & a_{23}Y_1^4\\
0 & a_{23}Y_1^4 & a_{33}Y_{1'}^4
\end{pmatrix} \) &

\( v_L\begin{pmatrix}
a_{11}Y_1^4 & 0 & a_{13}Y_{1'}^4\\
0 & a_{22}Y_{1'}^4 & a_{23}Y_1^4\\
a_{13}Y_{1'}^4 & a_{23}Y_1^4 & 0
\end{pmatrix} \) &

\( v_L\begin{pmatrix}
a_{11}Y_1^8 & 0 & a_{13}Y_{1''}^8\\
0 & 0 & a_{23}Y_1^6\\
a_{13}Y_{1''}^8 & a_{23}Y_1^6 & a_{33}Y_{1'}^8
\end{pmatrix} \) &

\( v_L\begin{pmatrix}
a_{11}Y_1^8 & a_{12}Y_{1''}^8 & 0\\
a_{12}Y_{1''}^8 & a_{22}Y_{1'}^8 & a_{23}Y_1^6\\
0 & a_{23}Y_1^6 & 0
\end{pmatrix} \) \\
\hline
\end{tabular}
}
\caption{Mass matrix \( M_L \) for textures \( B_1 \), \( B_2 \), \( B_3 \), and \( B_4 \)}
\label{W3T5}
\end{table}
The $A_{4}$-invariant superpotentials for the right-handed neutrinos corresponding to textures $B_{1}$, $B_{2}$, $B_{3}$, and $B_{4}$ are given in equations~\eqref{W3Q12a}, \eqref{W3Q12b}, \eqref{W3Q12c}, and \eqref{W3Q12d}, respectively.
\begin{subequations}\label{W3Q12}
\begin{align}
  \mathcal{W}^{(B_{1})}_{R} =\ & a_{11} \Delta^{c}_{R} L^{c^T}_{R_{1}} Y^{4}_{1} L^{c}_{R_{1}} 
  + a_{12} (\Delta^{c}_{R} L^{c^T}_{R_{1}} Y^{4}_{1^{\prime}} L^{c}_{R_{2}} + \Delta^{c}_{R} L^{c^T}_{R_{2}} Y^{4}_{1^{\prime}} L_{R_{1}}) \notag\\
  & + a_{23} (\Delta^{c}_{R} L^{c^T}_{R_{2}} Y^{4}_{1} L^{c}_{R_{3}} + \Delta^{c}_{R} L^{c^T}_{R_{3}} Y^{4}_{1} L^{c}_{R_{2}}) 
  + a_{33} \Delta^{c}_{R} L^{c^T}_{R_{3}} Y^{4}_{1^{\prime}} L^{c}_{R_{3}} \label{W3Q12a} \\[1ex]
  \mathcal{W}^{(B_{2})}_{R} =\ & a_{11} \Delta^{c}_{R} L^{c^T}_{R_{1}} Y^{4}_{1} L^{c}_{R_{1}} 
  + a_{13} (\Delta^{c}_{R} L^{c^T}_{R_{1}} Y^{4}_{1^{\prime}} L^{c}_{R_{3}} + \Delta^{c}_{R} L^{c^T}_{R_{3}} Y^{4}_{1^{\prime}} L^{c}_{R_{1}}) \notag\\
  & + a_{22} \Delta^{c}_{R} L^{c^T}_{R_{2}} Y^{4}_{1} L^{c}_{R_{2}} 
  + a_{23} (\Delta^{c}_{R} L^{c^T}_{R_{2}} Y^{4}_{1} L^{c^T}_{R_{3}} + \Delta^{c}_{R} L^{c^T}_{R_{3}} Y^{4}_{1} L^{c}_{R_{2}}) \label{W3Q12b} \\[4ex]
  \mathcal{W}^{(B_{3})}_{R} =\ & a_{11} \Delta^{c}_{R} L^{c^T}_{R_{1}} Y^{8}_{1} L^{c}_{R_{1}} 
  + a_{13} (\Delta^{c}_{R} L^{c^T}_{R_{1}} Y^{8}_{1^{\prime\prime}} L^{c}_{R_{3}} + \Delta^{c}_{R} L^{c^T}_{R_{3}} Y^{8}_{1^{\prime\prime}} L^{c}_{R_{1}}) \notag\\
  & + a_{23} (\Delta^{c}_{R} L^{c^T}_{R_{2}} Y^{6}_{1} L^{c}_{R_{3}} + \Delta^{c}_{R} L^{c^T}_{R_{3}} Y^{6}_{1} L^{c}_{R_{2}}) 
  + a_{33} \Delta^{c}_{R} L^{c^T}_{R_{2}} Y^{8}_{1^{\prime}} L^{c}_{R_{2}} \label{W3Q12c} \\[1ex]
  \mathcal{W}^{(B_{4})}_{R} =\ & a_{11} \Delta^{c}_{R} L^{c^T}_{R_{1}} Y^{8}_{1} L^{c}_{R_{1}} 
  + a_{12} (\Delta^{c}_{R} L^{c^T}_{R_{1}} Y^{8}_{1^{\prime\prime}} L^{c}_{R_{2}} + \Delta^{c}_{R} L^{c^T}_{R_{2}} Y^{4}_{1^{\prime\prime}} L^{c}_{R_{1}}) \notag\\
  & + a_{22} \Delta^{c}_{R} L^{c^T}_{R_{2}} Y^{8}_{1^{\prime}} L^{c}_{R_{2}} 
  + a_{23} (\Delta^{c}_{R} L^{c^T}_{R_{2}} Y^{6}_{1} L^{c}_{R_{3}} + \Delta^{c}_{R} L^{c^T}_{R_{3}} Y^{6}_{1} L^{c}_{R_{2}}) \label{W3Q12d}
\end{align}
\end{subequations}
 The structure of $M_{R}$ is similar to that of $M_{L}$, and we can write $M_{R}$ as follows.
\begin{equation}
    M_{R}=\frac{v_{R}}{v_{L}}M_{L}
\end{equation}
Similarly, we can write the superpotential for the neutrino sterile mixing for all four textures in the following way
\begin{subequations}\label{W3Q13}
\begin{align}
    \mathcal{W}^{(B_{1})}_{NS} &= g_{1} \chi^{c}_{R} L^{c^T}_{R_{1}} (Y^{2}_{3}S)_{1} + g_{2} \chi^{c}_{R} L^{c^T}_{R_{2}} (Y^{2}_{3}S)_{1^{\prime}}+g_{3} \chi^{c}_{R} L^{c^T}_{R_{3}} (Y^{2}_{3}S)_{1^{\prime\prime}} \label{W3Q13a} \\
    \mathcal{W}^{(B_{2})}_{NS} &= g_{1} \chi^{c}_{R} L^{c^T}_{R_{1}} (Y^{2}_{3}S)_{1} + g_{2} \chi^{c}_{R} L^{c^T}_{R_{2}} (Y^{2}_{3}S)_{1^{\prime\prime}}+g_{3} \chi^{c}_{R} L^{c^T}_{R_{3}} (Y^{2}_{3}S)_{1^{\prime}}\label{W3Q13b} \\
    \mathcal{W}^{(B_{3})}_{NS} &=   g_{1} \chi^{c}_{R} L^{c^T}_{R_{1}} (Y^{4}_{3}S)_{1} + g_{2} \chi^{c}_{R} L^{c^T}_{R_{2}} (Y^{4}_{3}S)_{1^{\prime}}+g_{3} \chi^{c}_{R} L^{c^T}_{R_{3}} (Y^{4}_{3}S)_{1^{\prime\prime}} \label{W3Q13c} \\
    \mathcal{W}^{(B_{4})}_{NS} &=  g_{1} \chi^{c}_{R} L^{c^T}_{R_{1}} (Y^{4}_{3}S)_{1} + g_{2} \chi^{c}_{R} L^{c^T}_{R_{2}} (Y^{4}_{3}S)_{1^{\prime\prime}}+g_{3} \chi^{c}_{R} L^{c^T}_{R_{3}} (Y^{4}_{3}S)_{1^{\prime}}  \label{W3Q13d}
\end{align}
\end{subequations}
The mixing matrices obtained for textures $B_{1}$, $B_{2}$, $B_{3}$, and $B_{4}$ from the superpotential given in equation~\eqref{W3Q13} are presented in Table~\ref{W3T6}.
\begin{table}[htbp]
\centering
\renewcommand{\arraystretch}{2.0}
\resizebox{\textwidth}{!}{
\begin{tabular}{|c|c|c|c|c|}
\hline
\textbf{Matrix} & \textbf{\( B_1 \)} & \textbf{\( B_2 \)} & \textbf{\( B_3 \)} & \textbf{\( B_4 \)} \\
\hline
\( M \) & 
\( v'\begin{pmatrix}
g_1Y_1 & g_1Y_3 & g_1Y_2\\
g_2Y_2 & g_2Y_1 & g_2Y_3\\
g_3Y_3 & g_3Y_2 & g_3Y_1
\end{pmatrix} \) &

\( v'\begin{pmatrix}
g_1Y_1 & g_1Y_3 & g_1Y_2\\
g_2Y_3 & g_2Y_2 & g_2Y_1\\
g_3Y_2 & g_3Y_1 & g_3Y_3
\end{pmatrix} \) &

\( v'\begin{pmatrix}
g_1(Y_1^2 - Y_2Y_3) & g_1(Y_2^2 - Y_1Y_3) & g_1(Y_3^2 - Y_1Y_2)\\
g_2(Y_3^2 - Y_1Y_2) & g_2(Y_1^2 - Y_2Y_3) & g_2(Y_2^2 - Y_1Y_2)\\
g_3Y_3 & g_3Y_2 & g_3Y_1
\end{pmatrix} \) &

\( v'\begin{pmatrix}
g_1(Y_1^2 - Y_2Y_3) & g_1(Y_2^2 - Y_1Y_3) & g_1(Y_3^2 - Y_1Y_2)\\
g_2(Y_3^2 - Y_1Y_2) & g_2(Y_1^2 - Y_2Y_3) & g_2(Y_2^2 - Y_1Y_2)\\
g_3Y_2 & g_3Y_1 & g_3Y_3
\end{pmatrix} \) \\
\hline
\end{tabular}
}
\caption{Mass matrix \( M \) for textures \( B_1 \), \( B_2 \), \( B_3 \), and \( B_4 \)}
\label{W3T6}
\end{table}
\subsection{Texture $A_{1}$ and $A_{2}$}
The charge assignments and modular weights considered to realize the $A_{1}$ and $A_{2}$ textures of the neutrino mass matrix are given in Table \ref{W3T7}.
\begin{table}[ht]
\centering
\resizebox{\textwidth}{!}{
\begin{tabular}{|p{2.2cm}|c|c|c|c|c|c|c|c|}
\hline
\textbf{Field} & $L^{c}_{R_{i}}$ & $L_{L_{i}}$ & $S$ & $\Phi$ & $\Delta^{c}_{R}$ & $\chi^{c}_{R}$ & $\Delta_{L}$ & $\chi_{L}$ \\ \hline

\multicolumn{9}{|c|}{\textbf{Texture $A_{1}$}} \\ \hline
$A_{4}$ & $1^{\prime},1,1^{\prime\prime}$ & $1^{\prime\prime},1,1^{\prime}$ & $3$ & $1$ & $1$ & $1$ & $1$ & $1$ \\
$k_{I}$ & $-2,0,-2$ & $-2,-4,-4$ & 0 & 0 & 0 & 0 & 0 & 0 \\ \hline

\multicolumn{9}{|c|}{\textbf{Texture $A_{2}$}} \\ \hline
$A_{4}$ & $1^{\prime},1^{\prime\prime},1$ & $1^{\prime\prime},1^{\prime},1$  & $3$ & $1$ & $1$ & $1$ & $1$ & $1$ \\
$k_{I}$ & $-2,-2,0$ & $-2,-4,-4$ & 0 & 0 & 0 & 0 & 0 & 0 \\ \hline

\end{tabular}}
\caption{Charge assignments and modular weights under $A_{4}$ for the particle content of the model for textures $A_{1}$, and $A_{2}$.}
\label{W3T7}
\end{table}
The $A_{4}$ invariant superpotential for the left-handed neutrino interaction and the right-handed neutrino interaction are given in equations \eqref{W3Q14} and \eqref{W3Q15}, respectively. The resulting mass matrix is presented in Table \ref{W3T8}.
\begin{subequations}\label{W3Q14}
\begin{align}
  \mathcal{W}^{(A_{1})}_{L} =\ & a_{13} (\Delta_{L} L^{T}_{L_{1}} Y^{6}_{1} L_{L_{3}} + \Delta_{L} L^{T}_{L_{3}} Y^{6}_{1} L_{L_{1}}) + a_{22}\Delta_{L}L^{T}_{L_{2}}Y^{8}_{1}L_{L_{2}} \notag\\
  & + a_{23} (\Delta_{L} L^{T}_{L_{2}} Y^{8}_{1^{\prime\prime}} L_{L_{3}} + \Delta_{L} L^{T}_{L_{3}} Y^{8}_{1^{\prime\prime}} L_{L_{2}}) 
  + a_{33} \Delta_{L} L^{T}_{L_{3}} Y^{8}_{1^{\prime}} L_{L_{3}} \label{W3Q14a} \\
  \mathcal{W}^{(A_{2})}_{L} =\ &  a_{12} (\Delta_{L} L^{T}_{L_{1}} Y^{6}_{1} L_{L_{2}} + \Delta_{L} L^{T}_{L_{2}} Y^{6}_{1} L_{L_{1}}) + a_{22} \Delta_{L} L^{T}_{L_{2}} Y^{8}_{1^{\prime}} L_{L_{2}} \notag\\
  & + a_{23} (\Delta_{L} L^{T}_{L_{2}} Y^{8}_{1^{\prime\prime}} L_{L_{3}} + \Delta_{L} L^{T}_{L_{3}} Y^{8}_{1^{\prime\prime}} L_{L_{2}})+a_{22}\Delta_{L}L^{T}_{L_{3}}Y^{8}_{1}L_{L_{3}} \label{W3Q14b} 
\end{align}
\end{subequations}
\begin{subequations}\label{W3Q15}
\begin{align}
  \mathcal{W}^{(A_{1})}_{R} =\ & a_{13} (\Delta^{c}_{R} L^{c^T}_{R_{1}} Y^{6}_{1} L^{c}_{R_{3}} + \Delta^{c}_{R} L^{c^T}_{R_{3}} Y^{6}_{1} L^{c}_{R_{1}}) + a_{22}\Delta_{R}L^{c^T}_{R_{2}}Y^{8}_{1}L^{c}_{R_{2}} \notag\\
  & + a_{23} (\Delta^{c}_{R} L^{c^T}_{R_{2}} Y^{8}_{1^{\prime\prime}} L^{c}_{R_{3}} + \Delta^{c}_{R} L^{c^T}_{R_{3}} Y^{8}_{1^{\prime\prime}} L^{c}_{R_{2}}) 
  + a_{33} \Delta^{c}_{R} L^{c^T}_{R_{3}} Y^{8}_{1^{\prime}} L^{c}_{R_{3}} \label{W3Q15a} \\[1ex]
  \mathcal{W}^{(A_{2})}_{R} =\ &  a_{12} (\Delta^{c}_{R} L^{c^T}_{R_{1}} Y^{6}_{1} L^{c}_{R_{2}} + \Delta^{c}_{R} L^{c^T}_{R_{2}} Y^{6}_{1} L^{c}_{R_{1}}) + a_{22} \Delta^{c}_{R} L^{c^T}_{R_{2}} Y^{8}_{1^{\prime}} L^{c}_{R_{2}} \notag\\
  & + a_{23} (\Delta^{c}_{R} L^{c^T}_{R_{2}} Y^{8}_{1^{\prime\prime}} L^{c}_{R_{3}} + \Delta^{c}_{R} L^{c^T}_{R_{3}} Y^{8}_{1^{\prime\prime}} L^{c}_{R_{2}})+a_{22}\Delta^{c}_{R}L^{c^T}_{R_{3}}Y^{8}_{1}L^{c}_{R_{3}} \label{W3Q15b} 
\end{align}
\end{subequations}
\begin{table}[htbp]
\centering
\renewcommand{\arraystretch}{1.2}
\setlength{\tabcolsep}{10pt}
\begin{tabular}{|c|c|c|}
\hline
\textbf{Matrix} & \textbf{Texture $A_{1}$} & \textbf{Texture $A_{2}$} \\
\hline
$M_{L}$ & 
$ v_{L} \begin{pmatrix}
0 & 0 & a_{13}Y^{6}_{1} \\
0 &  a_{22}Y_{1}^{8} & a_{23}Y^{8}_{1^{\prime\prime}} \\
a_{13}Y^{6}_{1} & a_{23}Y^{8}_{1^{\prime\prime}} & a_{33}Y^{8}_{1^{\prime}}
\end{pmatrix} $ &
$ v_{L} \begin{pmatrix}
0 & a_{12}Y^{6}_{1} & 0 \\
a_{12}Y^{6}_{1} & a_{22}Y^{8}_{1^{\prime}} & a_{23}Y^{8}_{1^{\prime\prime}} \\
0 & a_{23}Y^{8}_{1^{\prime\prime}} & a_{33}Y^{8}_{1}
\end{pmatrix} $ \\
\hline
\end{tabular}
\caption{Mass matrix $M_{L}$ for textures $A_{1}$ and $A_{2}$}
\label{W3T8}
\end{table}
Finally, the superpotential for right-handed and sterile neutrino mixing is given in equation \eqref{W3Q16}, and the corresponding mixing matrix is presented in Table \ref{W3T9}.
\begin{subequations}\label{W3Q16}
\begin{align}
    \mathcal{W}^{(A_{1})}_{NS} &= g_{1} \chi^{c}_{R} L^{c^T}_{R_{1}} (Y^{2}_{3}S)_{1^{\prime}} + g_{2} \chi^{c}_{R} L^{c^T}_{R_{2}} (Y^{4}_{3}S)_{1}+g_{3} \chi^{c}_{R} L^{c^T}_{R_{3}} (Y^{4}_{3}S)_{1^{\prime\prime}} \label{W3Q16a} \\
    \mathcal{W}^{(A_{2})}_{NS} &= g_{1} \chi^{c}_{R} L^{c^T}_{R_{1}} (Y^{2}_{3}S)_{1^{\prime}} + g_{2} \chi^{c}_{R} L^{c^T}_{R_{2}} (Y^{4}_{3}S)_{1^{\prime\prime}}+g_{3} \chi^{c}_{R} L^{c^T}_{R_{3}} (Y^{4}_{3}S)_{1}\label{W3Q16b}
\end{align}
\end{subequations}

\begin{table}[htbp]
\centering
\renewcommand{\arraystretch}{1.7}
\setlength{\tabcolsep}{12pt}
\resizebox{\textwidth}{!}{
\begin{tabular}{|c|c|c|}
\hline
\textbf{Matrix} & \textbf{Texture $A_{1}$} & \textbf{Texture $A_{2}$} \\
\hline
$M$ & 
$ v^{\prime} \begin{pmatrix}
g_{1}Y_{2} & g_{1}Y_{1} & g_{1}Y_{3} \\
g_{2}(Y^{2}_{1}-Y_{2}Y_{3}) & g_{2}(Y^{2}_{2}-Y_{1}Y_{3}) & g_{2}(Y^{2}_{3}-Y_{1}Y_{2}) \\
g_{3}(Y^{2}_{2}-Y_{1}Y_{3}) & g_{3}(Y^{2}_{3}-Y_{1}Y_{2}) & g_{3}(Y^{2}_{1}-Y_{2}Y_{3})
\end{pmatrix} $ &
$ v^{\prime} \begin{pmatrix}
g_{1}Y_{2} & g_{1}Y_{1} & g_{1}Y_{3} \\
g_{2}(Y^{2}_{2}-Y_{1}Y_{3}) & g_{2}(Y^{2}_{3}-Y_{1}Y_{2}) & g_{2}(Y^{2}_{1}-Y_{2}Y_{3}) \\
 g_{3}(Y^{2}_{1}-Y_{2}Y_{3})& g_{3}(Y^{2}_{2}-Y_{1}Y_{3}) & g_{3}(Y^{2}_{3}-Y_{1}Y_{2})
\end{pmatrix} $ \\
\hline
\end{tabular} }
\caption{Mass matrix $M$ for textures $A_{1}$ and $A_{2}$}
\label{W3T9}
\end{table}
\section{Experimental Details}\label{s4}
\subsection{DUNE}
The Deep Underground Neutrino Experiment (DUNE)\cite{Marciano:2006uc,Bass:2013vcg, DUNE:2015lol,DUNE:2016hlj} is a long-baseline neutrino oscillation experiment designed primarily to determine the neutrino mass hierarchy and the CP-violating phase. It also aims to detect low-energy neutrino events and search for proton decay or other beyond Standard Model phenomena. The setup consists of two detectors: a near detector, which will be placed at Fermilab, 574 m downstream of the neutrino production point, and a large liquid argon time-projection chamber (LArTPC) far detector (FD) located at the 4850 ft level of the Sanford Underground Research Facility (SURF) in Lead, South Dakota, 1285 km from the neutrino production point. 

The neutrinos are produced using a 120 GeV proton beam with a beam power of 1.2 MW from Fermilab's Main Injector, which is directed onto a graphite target, followed by a horn-focusing system. To address questions such as the neutrino mass hierarchy, the value of the CP-violating phase $\delta_{\text{CP}}$, and the octant of the atmospheric mixing angle, DUNE investigates the $\nu_{\mu} \rightarrow \nu_{e}$ ($\bar{\nu}_{\mu} \rightarrow \bar{\nu}_{e}$) oscillation channels and studies the energy dependence of the $\nu_{e}$ ($\bar{\nu}_{e}$) appearance probability.
\subsection{T2HK}
T2HK (Tokai to Hyper-Kamiokande)\cite{Hyper-KamiokandeProto-:2015xww} is a proposed next-generation long-baseline neutrino oscillation experiment designed to achieve high-precision measurements of leptonic CP violation, determine the neutrino mass ordering, and improve constraints on neutrino mixing parameters. The experiment features a baseline of 295~km, with an intense neutrino beam produced at the J-PARC accelerator complex and detected at the Hyper-Kamiokande (HK) detector. In the proposed setup, the J-PARC neutrino beam will operate at a beam power of 1.3~MW, corresponding to an annual exposure of $27 \times 10^{21}$ protons on target. The beamline employs the off-axis technique, placing the detector at an angle of $2.5^\circ$ relative to the beam direction.

\section{\textbf{Numerical Analysis}}\label{s5}
While constructing the model, we have used free parameters, all of which are considered to be complex numbers. The vacuum expectation value (VEV) of the bidoublet $\Phi$ is taken to be 246 GeV. The VEV of the left-handed scalar triplet is considered as $v_{L} = 0.01$ eV, while the VEV of the right-handed scalar triplet and doublet is taken to be approximately 10 TeV.\\
We have realized different textures of the neutrino mass matrix using the $\Gamma_3$ modular group. Specifically, we have constructed the seven two-zero textures in terms of modular forms. The active neutrino mass matrix can be diagonalized by the $U_{PMNS}$ matrix, and can therefore be written as:
\begin{equation}\label{W3Q17}
    m_{\text{diag}} = U_{PMNS} \, m_{\nu} \, U_{PMNS}^T
\end{equation}
where $m_{\text{diag}} = \text{diag}(m_1, m_2, m_3)$. Neutrino oscillation experiments determine two mass-squared differences, three mixing angles, and the Dirac CP-violating phase $\delta_{CP}$. Assuming the lightest neutrino mass lies in the range $10^{-5}$--$0.1~\text{eV}$, the remaining two mass eigenvalues can be expressed in terms of the measured mass-squared differences as follows.
\begin{equation}\label{W3Q18}
\begin{aligned}
    m_{\text{diag}} &= \left(m_1, \sqrt{m_1^2 + \Delta m^2_{21}}, \sqrt{m_1^2 + \Delta m^2_{31}}\right) \quad &\text{(for NH)} \\
    m_{\text{diag}} &= \left(\sqrt{m_3^2 + \Delta m^2_{23} - \Delta m^2_{21}}, \sqrt{m_3^2 + \Delta m^2_{23}}, m_3\right) \quad &\text{(for IH)}
\end{aligned}
\end{equation}
Using equation~\eqref{W3Q17} along with the $3\sigma$ values of the oscillation parameters, we can compute the active neutrino mass matrix $m_{\nu}$ in terms of known quantities. Since we have already expressed $m_{\nu}$ in terms of Yukawa couplings for the seven two-zero textures, we equate the two forms of $m_{\nu}$ to solve for the unknown Yukawa couplings in each case. Once the Yukawa couplings are determined, we calculate the eigenvalues and eigenvectors of the resulting active neutrino mass matrices to calculate the neutrino masses and oscillation parameters corresponding to each realized texture for both NH and IH. The mixing angles can be computed using the following relations:
\begin{align}\label{W3Q19}
\sin^{2}\theta_{13} = |(U_{\nu})_{13}|^{2} ~,~ \sin^{2}\theta_{23} = \frac{|(U_{\nu})_{23}|^{2}}{1-|(U_{\nu})_{13}|^{2}} ~,~\sin^{2}\theta_{12}=\frac{|(U_{\nu})_{12}|^{2}}{1-|(U_{\nu})_{13}|^{2}}
\end{align}
The Dirac CP phase $\delta_{CP}$, Jarlskog invariant $J_{CP}$, and Majorana phases $\alpha$, $\beta$ can also be extracted from the $U_{\nu}$ matrix using the following expressions:
\begin{equation}\label{W3Q20}
    J_{CP} = \text{Im}[U_{e1} U_{\mu2} U_{e2}^* U_{\mu1}^*] = s_{23} c_{23} s_{12} c_{12} s_{13} c_{13}^2 \sin\delta_{CP}
\end{equation}
\begin{equation}\label{W3Q21}
    \text{Im}[U_{e1}^* U_{e2}] = c_{12} s_{12} c_{13}^2 \sin\alpha~, \quad
    \text{Im}[U_{e1}^* U_{e3}] = c_{12} s_{13} c_{13} \sin(\beta - \delta_{CP})
\end{equation}
We have calculated the effective Majorana mass $m_{\text{eff}}$ for each texture and found that, for the five textures with a non-zero $(1,1)$ element in the neutrino mass matrix, the predicted values of $m_{\text{eff}}$ are well below the experimental upper bounds. Furthermore, we computed the branching ratios for three lepton flavor violating (LFV) decay processes: $\mu \rightarrow e\gamma$, $\tau \rightarrow \mu\gamma$, and $\tau \rightarrow e\gamma$. In all cases, the predicted branching ratios for all textures lie well below the current experimental limits. The plots of $m_{\text{eff}}$ versus the lightest neutrino mass, along with the branching ratios versus the lightest neutrino mass, indicate that these observables can provide a suitable lower bound on the lightest neutrino mass. The formulas used to compute these quantities are discussed in \cite{Kumar:2024uxn,Kumar:2025bfe}, and the results obtained for different textures are summarised in Table~\ref{W3T11}.
In the numerical analysis section, we present the results and plots corresponding to textures $A_{1}$, $B_{4}$ and C. The results and plots for the remaining four textures are discussed in Appendix \ref{W3A4} and \ref{W3A5} 

\subsection{$\chi^{2}$ analysis}
Within the model framework, the light neutrino mass matrix was formulated by introducing the Yukawa couplings as free parameters. Once these couplings were determined, their values were incorporated into the mass matrix. The diagonalization was carried out by calculating its eigenvectors, which led to the construction of the unitary matrix responsible for the diagonalization. This unitary matrix was subsequently employed to derive the neutrino oscillation parameters. We feed these data in GLoBES to compute the sensitivities of these textures and hence to see how DUNE and DUNE+T2HK can constrain the $\theta_{23}-\delta_{CP} $ parameter space predicted by these textures. $\Delta\chi^{2}$ used in this work is defined as:
\begin{equation}
\begin{split}
    \Delta\chi^{2}(p^{\text{true}}) = \min_{p^{\text{test}},\eta} \Bigg[ & 2\sum_{i,j,k}^{} \left\{ N_{ijk}^{\text{test}}(p^{\text{test}};\eta) - N_{ijk}^{\text{true}}(p^{\text{true}}) + N_{ijk}^{\text{true}}(p^{\text{true}}) \ln\frac{N_{ijk}^{\text{true}}(p^{\text{true}})}{N_{ijk}^{\text{test}}(p^{\text{test}};\eta)} \right\} \\
    & + \sum_{l} \frac{(p_{l}^{\text{true}} - p_{l}^{\text{test}})^{2}}{\sigma_{p_{l}}^{2}} + \sum_{m} \frac{\eta_{m}^{2}}{\sigma_{\eta m}^{2}} \Bigg]
\end{split}
\label{chi2-eq}
\end{equation}
where:
\begin{itemize}
    \item $N^{\text{true}}$ is the simulated event rate corresponding to the true values of the oscillation parameters (treated as `data') and $N^{\text{test}}$ denotes the events simulated in the test or `fit' as per the model predictions,
    \item $p^{\text{true}}$ is the true/test set od oscillation parameters,
    \item Index $i$, $j$ and $k$ runs over energy bins, oscillation channels and running mode (neutrino or antineutrino) respectively.
\end{itemize}

For DUNE, an energy bin width of $0.125~\mathrm{GeV}$ is adopted up to $8~\mathrm{GeV}$.
For energies above $8~\mathrm{GeV}$ and extending to $110~\mathrm{GeV}$, variable bin
widths of $1~\mathrm{GeV}$, $2~\mathrm{GeV}$, and $10~\mathrm{GeV}$ are used.
For T2HK, the analysis employs $24$ uniform energy bins of width $0.05~\mathrm{GeV}$
covering the range $0.075$--$1.275~\mathrm{GeV}$.
  \\

The nuisance parameters $\eta_{m}$ in equation \ref{chi2-eq}, describe how the predicted event rates depend on various sources of systematic error. The full list of systematics and their corresponding nuisance parameters is provided in Table~\ref{tab:uncertainity}.

\begin{table}[H]
    \centering
    \renewcommand{\arraystretch}{1.0}
    \begin{tabular}{|c|c|c|c|}

        \hline
        \multirow{2}{*}{Experiment details} &\multirow{2}{*}{Channels} & \multicolumn{2}{|c|}{Normalization uncertainty} \\ \cline{3-4}
        & & Signal & Background \\
        \hline \hline
         \textbf{DUNE}, Baseline: 1300 km  &  & &\\
          &  $\nu_e (\bar \nu_e)$ appearance & 2\% (2\%) & 5\% (5\%) \\
          Fiducial mass = 40 kt (LArTPC) & $\nu_\mu (\bar \nu_\mu)$ disappearance & 5\% (5\%) & 5\% (5\%) \\ 
         \hline
    \textbf{T2HK}, Baseline: 295 km    &  & & \\
          &  $\nu_e (\bar \nu_e)$ appearance  & 3.2\% (3.9\%) & 5\% (5\%) \\ 
          Fiducial mass = 374 kt (WC)& $\nu_\mu (\bar \nu_\mu)$ disappearance  & 3.6\% (3.6\%) & 5\% (5\%)  \\ 
        \hline
    \end{tabular}
    \caption{List of uncertainties ($\sigma_{\eta}$) on the nuisance parameters ($\eta$) used in our simulation for DUNE and T2HK.}
    \label{tab:uncertainity}
\end{table}

In this work, we have explicitly shown the effect of the priors on $\sin^2\theta_{12}$ and $\sin^2\theta_{13}$ and the prior term used in equation ~\ref{chi2-eq} contributes to the total $\chi^2$ which is then minimized and $\chi^2_{min}$ is calculated. The precise measurements of $\theta_{12}$ from JUNO experiments has been incorporated as the prior on $\sin^2\theta_{12}$ and is defined as:

\begin{equation}
    \chi^2_{JUNO}(\sin^2{\theta_{12}})=(\frac{\sin^2{\theta_{12}}^{fit}-\sin^2{\theta_{12}}^{bf}}{0.0087})^{2}
\end{equation} 

Similarly using the values of 1$\sigma$ uncertainties on $\sin^2\theta_{13}$ from NuFit 6.0, we calculate the prior on $\sin^2\theta_{13}$ and hence minimized $\chi^2$ is being calculated. 

\begin{figure}[H]
  \centering
  \begin{subfigure}[b]{0.45\textwidth}
    \centering
    \includegraphics[width=\textwidth]{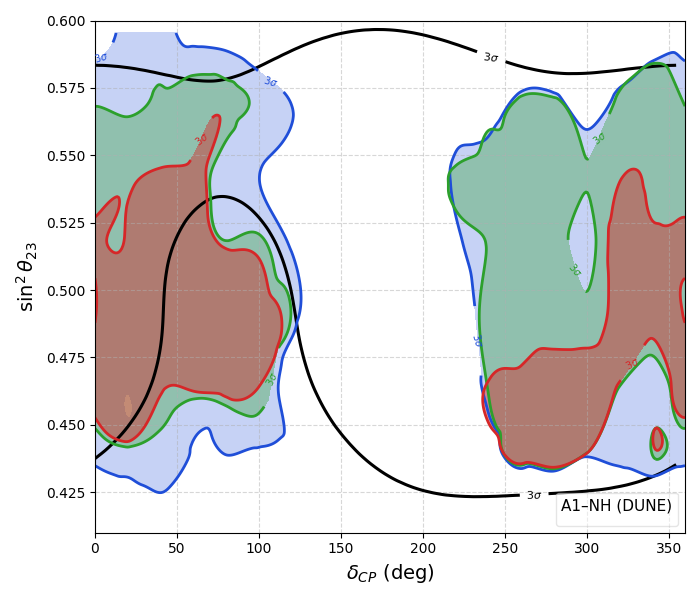}
    \label{fig:subA1}
  \end{subfigure}
  \begin{subfigure}[b]{0.45\textwidth}
    \centering
    \includegraphics[width=\textwidth]{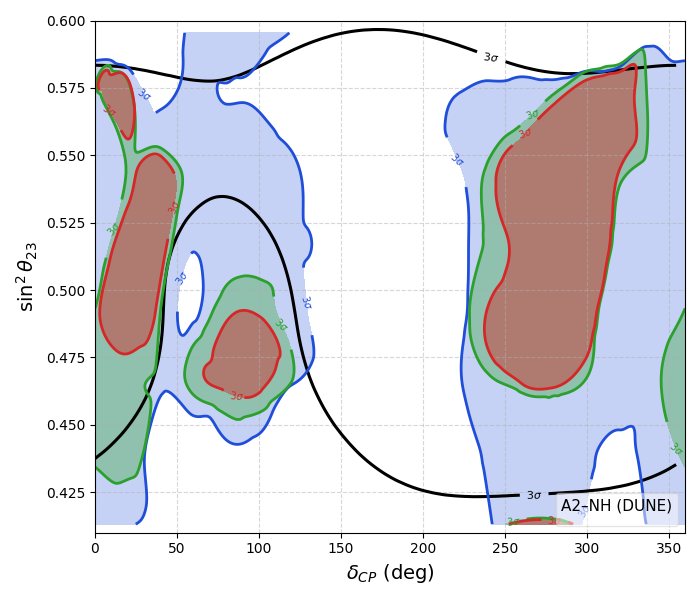}
    \label{fig:subB}
  \end{subfigure}
  \\[-9.5ex]  

  \begin{subfigure}[b]{0.45\textwidth}
    \centering
    \includegraphics[width=\textwidth]{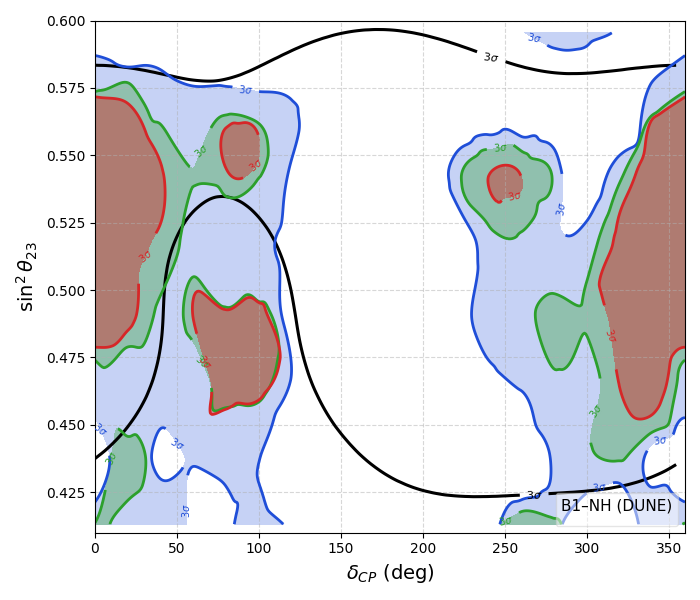}
    \label{fig:subC}
  \end{subfigure}
  \begin{subfigure}[b]{0.45\textwidth}
    \centering
    \includegraphics[width=\textwidth]{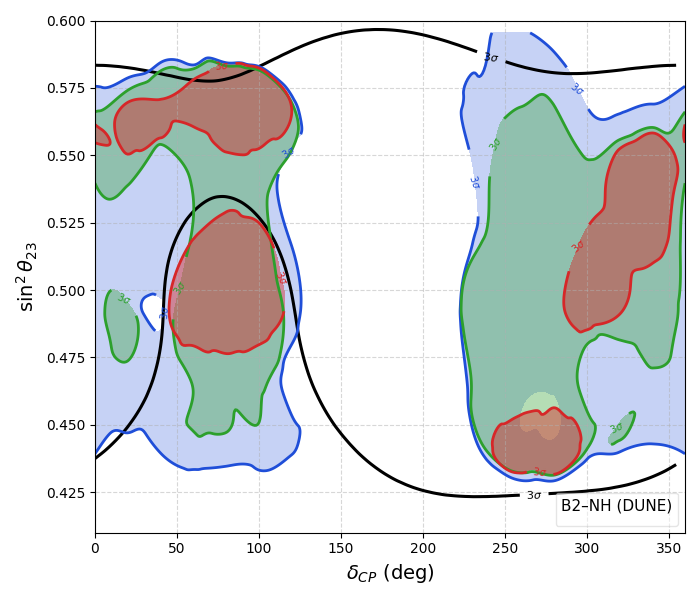}
    \label{fig:subD}
  \end{subfigure}
  \\[-9.5ex]
  \begin{subfigure}[b]{0.45\textwidth}
    \centering
    \includegraphics[width=\textwidth]{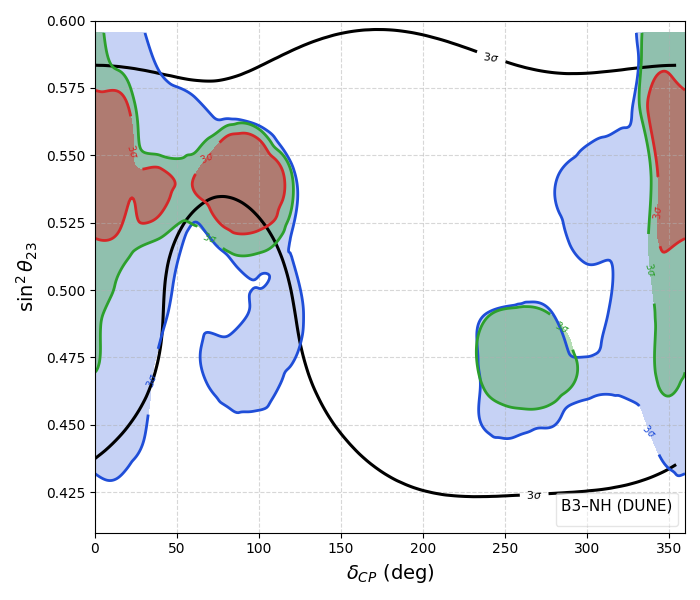}
    \label{fig:subC1}
  \end{subfigure}
  \begin{subfigure}[b]{0.45\textwidth}
    \centering
    \includegraphics[width=\textwidth]{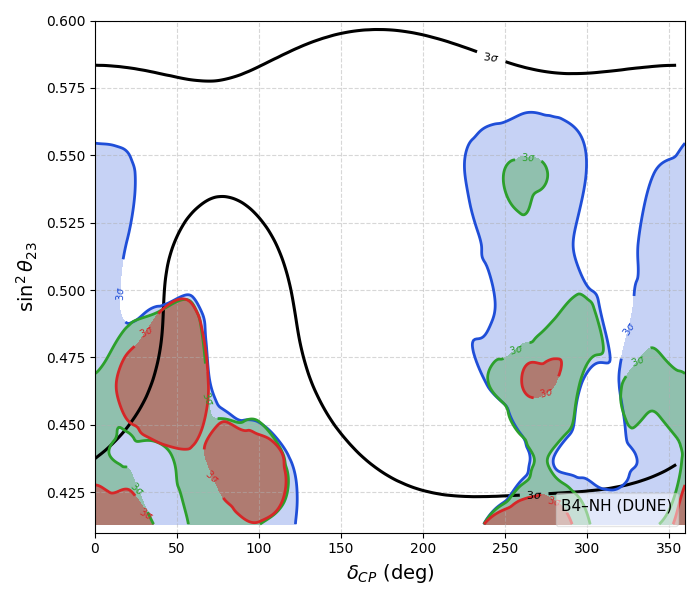}
    \label{fig:du}
  \end{subfigure}
  \\[-6ex]
  \caption{ 3$\sigma$ allowed regions in the $\theta_{23}$--$\delta_{\rm CP}$ plane for two-zero textures predicted by the modular $A_4$ left--right symmetric framework at DUNE in NH mode. The light blue band corresponds to the allowed region for different textures, while the light green (red) region shows the impact of including a prior on $\theta_{12}$ (both $\theta_{12}$ and $\theta_{13}$). The black line shows the globally allowed regions as per NuFit 6.0. when assumed hierarchy is normal.}
  \label{fig:NH}
\end{figure}

In this analysis, each texture hypothesis is tested at DUNE and DUNE+T2HK using oscillation parameters predicted by the texture-zero conditions. Assuming normal (inverted) mass ordering, we generate the data for $\theta_{12} = 33.68^{\circ}$, $\theta_{13} = 8.52^{\circ}$, $\Delta m^{2}_{21} = 7.49\times 10^{-5} \rm eV^{2}$, and $\Delta m^{2}_{31} = 2.534\times 10^{-3} \rm eV^{2}$ ($\Delta m^{2}_{32} = -2.51\times 10^{-3} \rm eV^{2}$).  THe true values of $\sin^2\theta_{23}$ and  $\delta_{CP}$ is varied over the full 3$\sigma$ allowed ranges to obtain the allowed parameter space for the different textures. 
 All six neutrino oscillation parameters are varied in the fit according to the constraints set by two-zero textures predicted by the modular $A_4$ left--right symmetric framework. \\

\subsection{Results of the $\chi^2$ analysis}
\begin{figure}[H]
  \centering
  \begin{subfigure}[b]{0.45\textwidth}
    \centering
    \includegraphics[width=\textwidth]{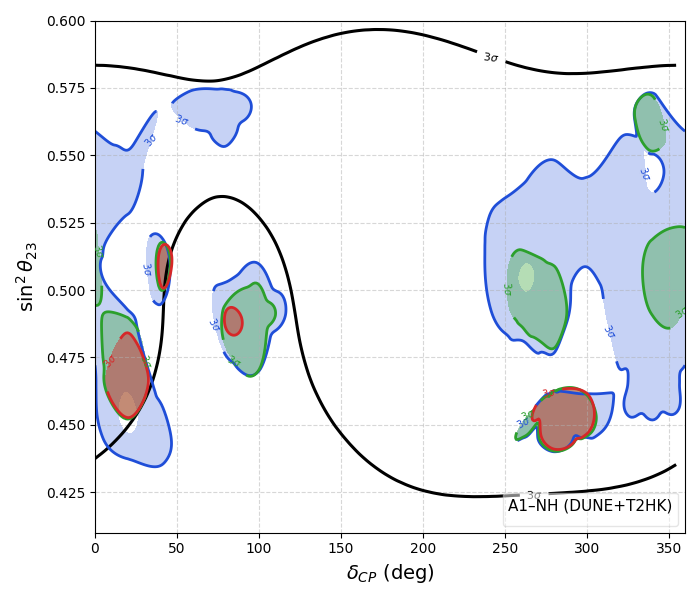}
    \label{fig:subA2}
  \end{subfigure}
  \begin{subfigure}[b]{0.45\textwidth}
    \centering
    \includegraphics[width=\textwidth]{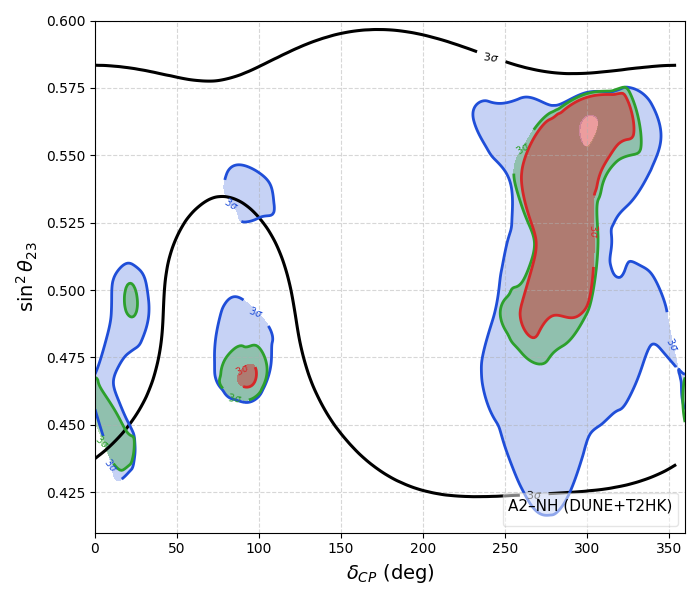}
    \label{fig:subB1}
  \end{subfigure}
  \\[-9.5ex]  

  \begin{subfigure}[b]{0.45\textwidth}
    \centering
    \includegraphics[width=\textwidth]{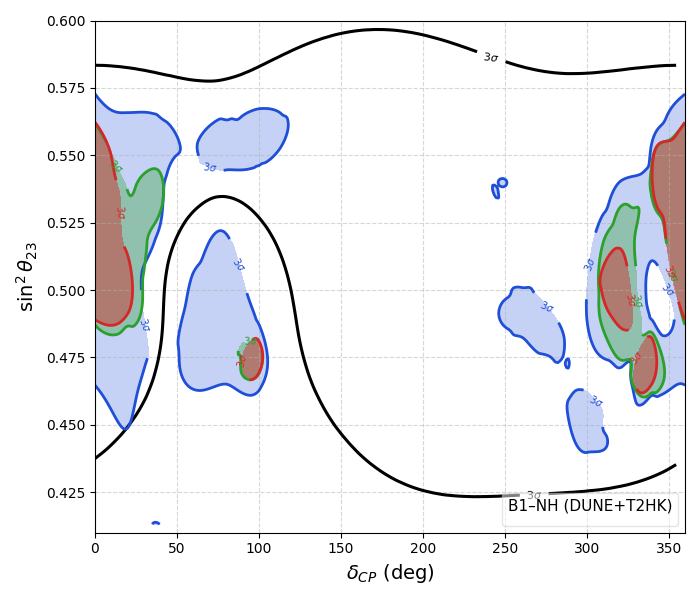}
    \label{fig:subC2}
  \end{subfigure}
  \begin{subfigure}[b]{0.45\textwidth}
    \centering
    \includegraphics[width=\textwidth]{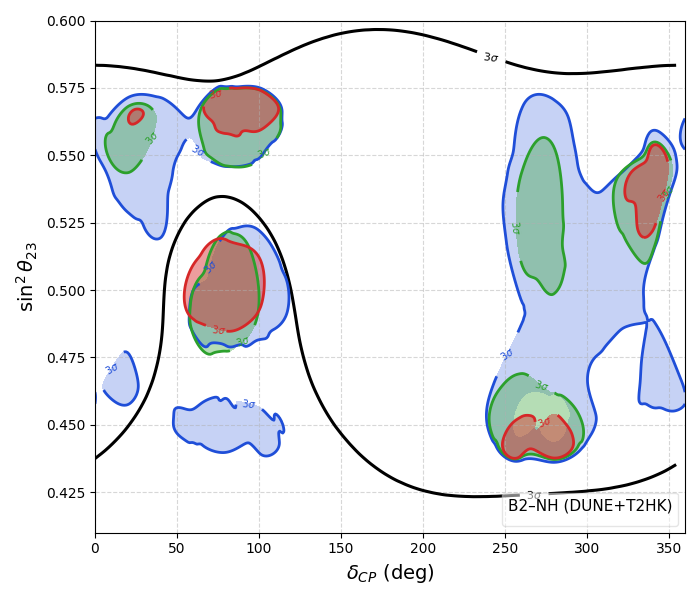}
    \label{fig:subD1}
  \end{subfigure}
  \\[-9.5ex]
    \begin{subfigure}[b]{0.45\textwidth}
    \centering
    \includegraphics[width=\textwidth]{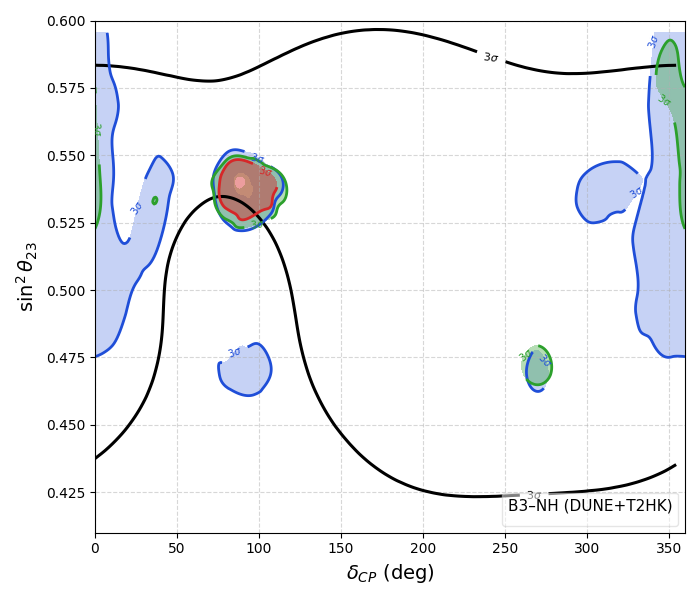}
    \label{fig:subC3}
  \end{subfigure}
  \begin{subfigure}[b]{0.45\textwidth}
    \centering
    \includegraphics[width=\textwidth]{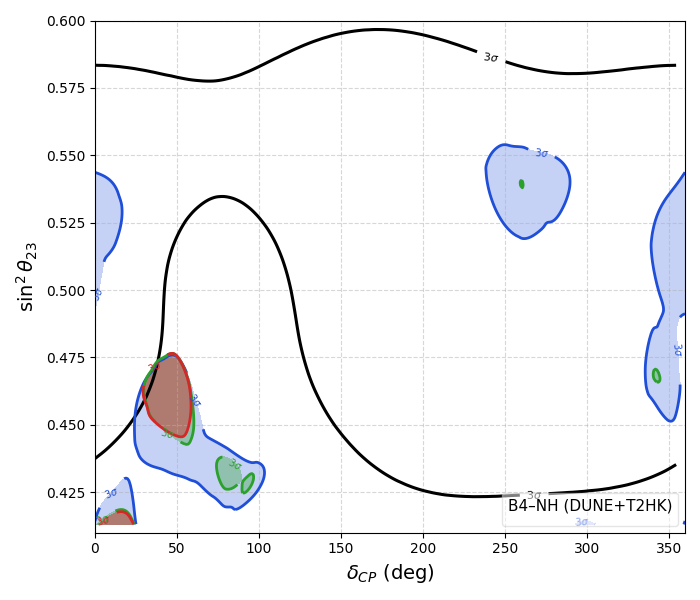}
    \label{fig:subD2}
  \end{subfigure}
  \\[-6ex]

  \caption{3$\sigma$ allowed regions in the $\theta_{23}$--$\delta_{\rm CP}$ plane for two-zero textures predicted by the modular $A_4$ left--right symmetric framework at DUNE+T2HK in NH mode. The colour code is same as Figure.~\ref{fig:NH}.}
  \label{fig:comb-NH}
\end{figure}

\begin{figure}[!h]
    \centering
    \includegraphics[width=0.46\textwidth]{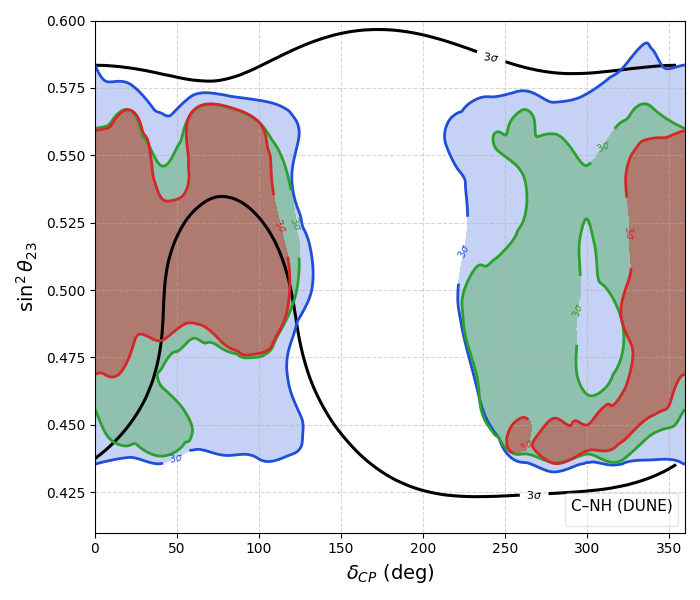}\hfill
    \includegraphics[width=0.46\textwidth]{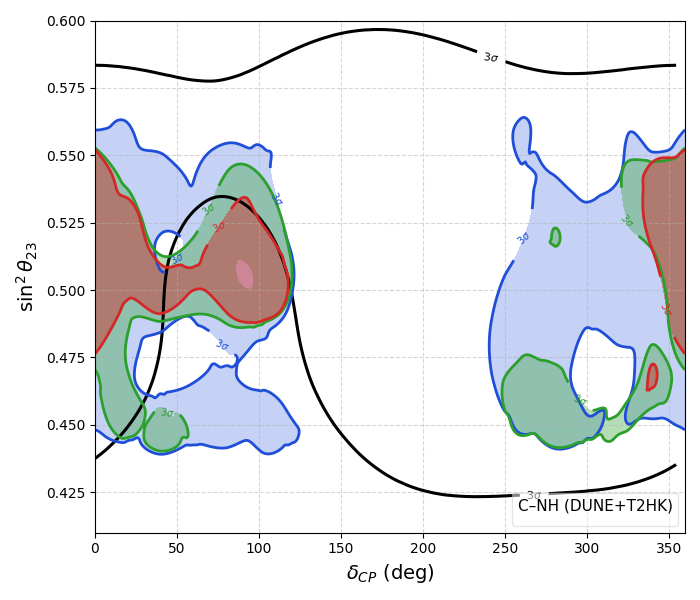}

    \caption{3$\sigma$ allowed regions in the $\theta_{23}$--$\delta_{\rm CP}$ plane for two-zero textures predicted by the modular $A_4$ left--right symmetric framework at DUNE (left) and DUNE+T2HK (right) when assumed mass ordering is normal. The colour code is same as Figure.~\ref{fig:NH}.}
    \label{c-nh}
\end{figure}

In Figs.~\ref{fig:NH}, \ref{fig:comb-NH}, and \ref{c-nh}, we present the allowed
regions in the $\theta_{23}$--$\delta_{\rm CP}$ (true) parameter space at DUNE and in the
combined DUNE+T2HK setup, assuming normal mass ordering as the true hierarchy. As
described in the previous section, each sub panel in Fig.~\ref{fig:NH} illustrates the
allowed parameter space for the $A_{1}$, $A_{2}$, and B-type ($B_{1}--B_{4}$) textures at DUNE, assuming
$13$ years of data taking, equally divided between neutrino and antineutrino modes.
 The light blue regions correspond to the parameter space allowed by
the texture-induced correlations alone and they span over multiple disjoint regions, mostly around the globally (black contour) allowed regions.
The inclusion of recent measurements from JUNO on $\theta_{12}$ as prior (light green
regions) leads to a noticeable reduction of the allowed parameter space for most textures.
It shows how the precision on $\theta_{12}$ plays an important role in
constraining the parameter space further. Adding priors on both $\theta_{12}$ and  $\theta_{13}$ further restricts the allowed regions (red) for all the textures and the allowed regions collapse into compact and well-localized islands in the $\theta_{23}$--$\delta_{\rm CP}$ plane. This demonstrates a significant enhancement in predictability of DUNE in presence of both the priors. \\

We now compare the results shown in Figs.~\ref{fig:NH} with \ref{fig:comb-NH}, where
Fig.~\ref{fig:comb-NH} presents the allowed parameter space obtained after combining
T2HK with DUNE. The inclusion of T2HK leads to a substantial reduction of the allowed
parameter space for almost all textures, as evident from Fig.~\ref{fig:comb-NH}. This
improvement arises from the complementarity of the two experiments, which helps to lift
degeneracies in the $\theta_{23}$--$\delta_{\rm CP}$ plane, particularly in the presence
of external priors.

For almost all the textures, the combined DUNE+T2HK analysis yields highly localized allowed regions
once priors on $\theta_{12}$ and $\theta_{13}$ are imposed. As a result, the parameter
space collapses into compact islands, many of which lie within the globally allowed
regions. At the same time, a significant fraction of the globally allowed parameter space
is no longer compatible with the texture predictions, indicating a notable enhancement in
predictivity compared to the DUNE-only scenarios. Similar pattern is observed in case of C two-zero texture as shown in fig.~\ref{c-nh}.

 This reduced overlap between the globally allowed region and the texture-predicted
parameter space in the combined DUNE+T2HK analysis demonstrates the potential of
future long-baseline experiments to test and discriminate among such predictions.

\begin{figure}[!h]
  \centering
  \begin{subfigure}[b]{0.45\textwidth}
    \centering
    \includegraphics[width=\textwidth]{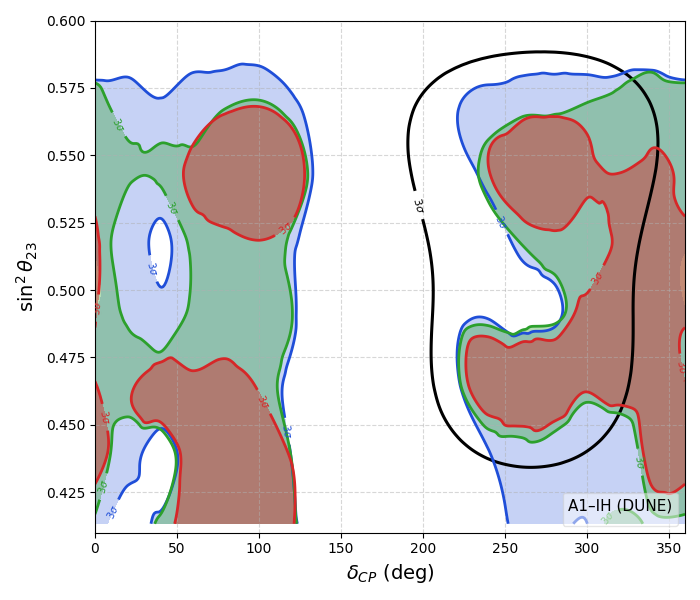}
    \label{fig:subA3}
  \end{subfigure}
  \begin{subfigure}[b]{0.45\textwidth}
    \centering
    \includegraphics[width=\textwidth]{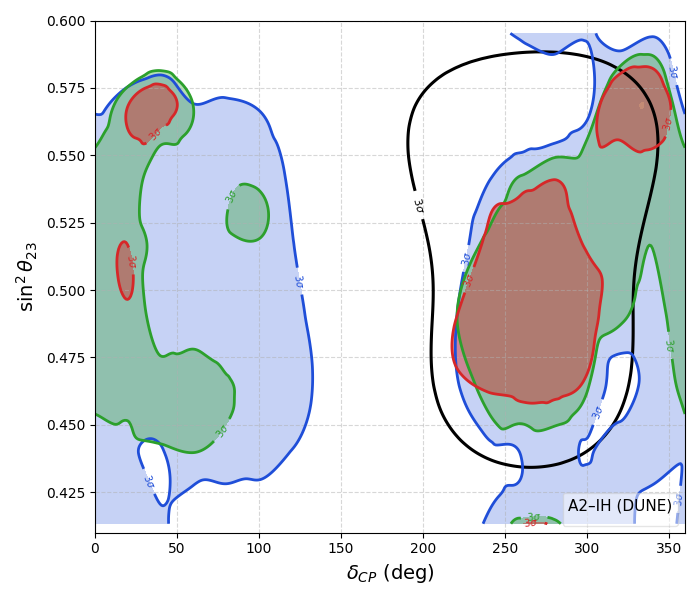}
    \label{fig:subB2}
  \end{subfigure}
  \\[-9ex]  

  \begin{subfigure}[b]{0.45\textwidth}
    \centering
    \includegraphics[width=\textwidth]{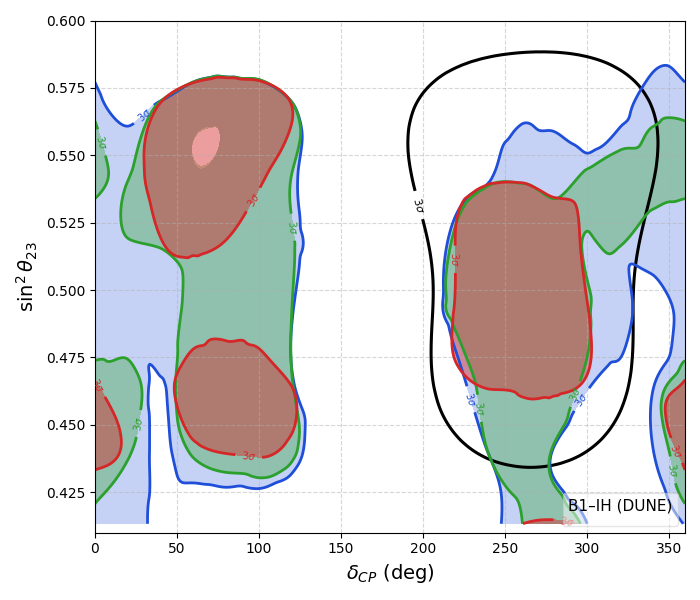}
    \label{fig:subC4}
  \end{subfigure}
  \begin{subfigure}[b]{0.45\textwidth}
    \centering
    \includegraphics[width=\textwidth]{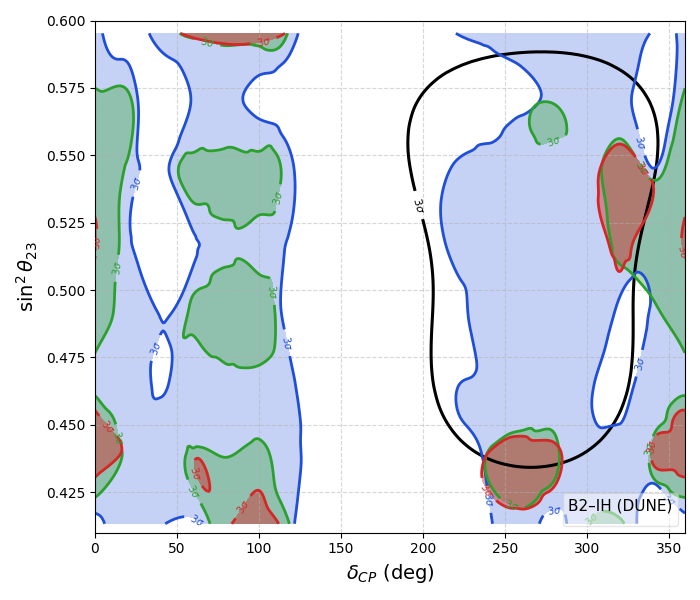}
    \label{fig:subD3}
  \end{subfigure}
  \\[-9ex]
    \begin{subfigure}[b]{0.45\textwidth}
    \centering
    \includegraphics[width=\textwidth]{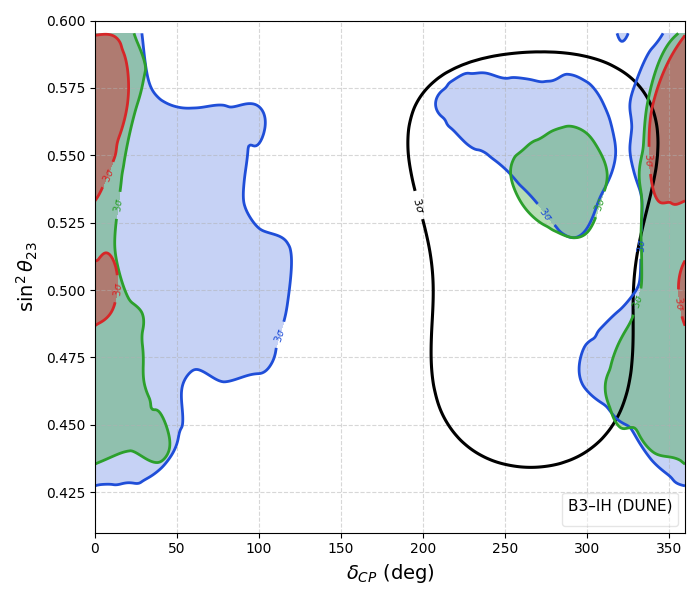}
    \label{fig:subC5}
  \end{subfigure}
  \begin{subfigure}[b]{0.45\textwidth}
    \centering
    \includegraphics[width=\textwidth]{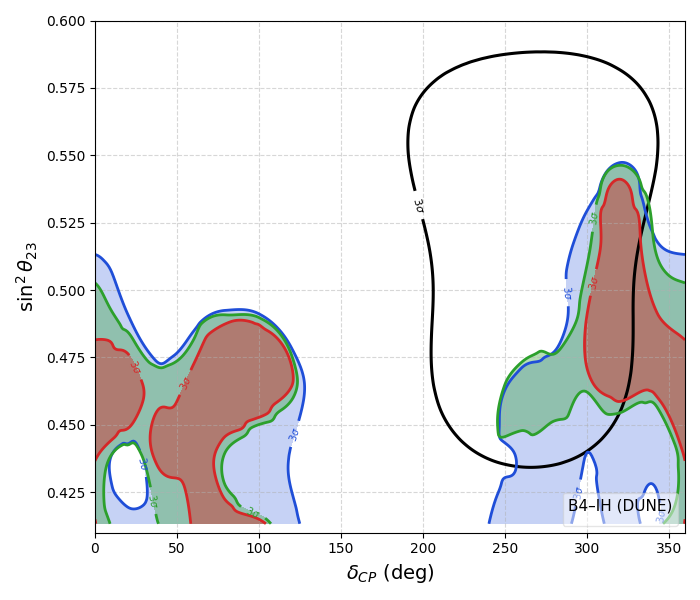}
    \label{fig:subD4}
  \end{subfigure}
  \\[-4ex]

  \caption{3$\sigma$ allowed regions in the $\theta_{23}$--$\delta_{\rm CP}$ plane for two-zero textures predicted by the modular $A_4$ left--right symmetric framework at DUNE when assumed mass ordering is inverted. The colour code is same as Figure.~\ref{fig:NH}.}
  \label{fig:IH-dune}
\end{figure}

Figures~\ref{fig:IH-dune} show the allowed regions in the
$\theta_{23}$--$\delta_{\rm CP}$ plane for different two-zero neutrino mass textures
predicted by the modular $A_4$ left--right symmetric framework at DUNE, assuming inverted
mass hierarchy as the true ordering. The colour coding follows the same convention as in
the normal hierarchy case. In comparison with the normal hierarchy cases, the predicted regions in IH mode are generally broader and more fragmented. But the inclusion of a prior on $\theta_{12}$ leads to a moderate reduction of the allowed
parameter space as shown by the light green regions. When priors on both $\theta_{12}$ and $\theta_{13}$ are imposed, the allowed regions (red)
shrink further. The allowed regions in $A_{2}$, $B_{1}$, $B_{2}$ and $B_{3}$ textures gets more localized in the
$\theta_{23}$--$\delta_{\rm CP}$ plane, indicating
enhancement predictability.

\begin{figure}[!h]
  \centering
  \begin{subfigure}[b]{0.45\textwidth}
    \centering
    \includegraphics[width=\textwidth]{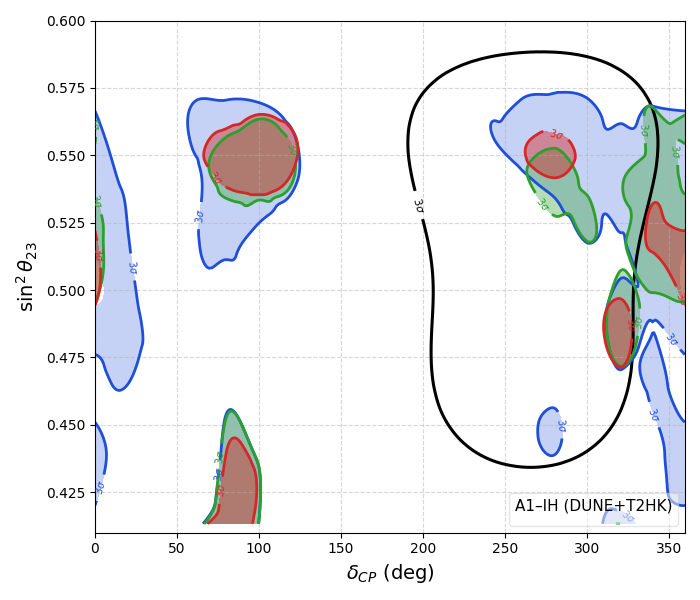}
    \label{fig:subA4}
  \end{subfigure}
  \begin{subfigure}[b]{0.45\textwidth}
    \centering
    \includegraphics[width=\textwidth]{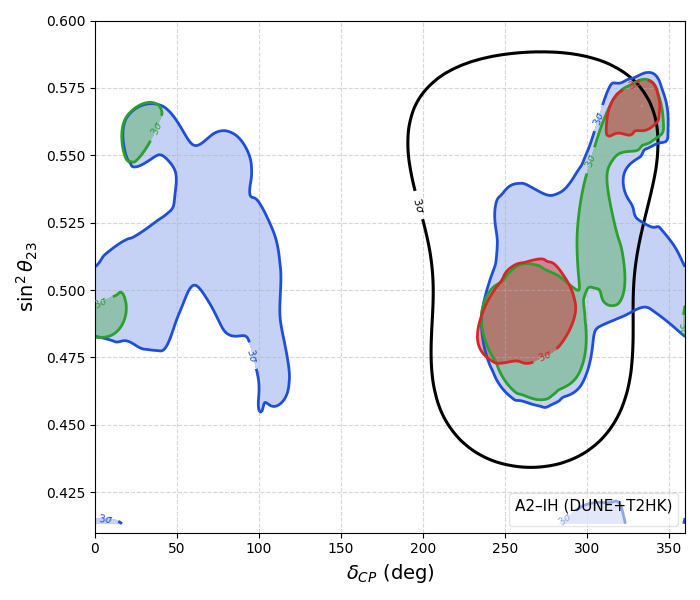}
    \label{fig:subB3}
  \end{subfigure}
  \\[-9ex]  

  \begin{subfigure}[b]{0.45\textwidth}
    \centering
    \includegraphics[width=\textwidth]{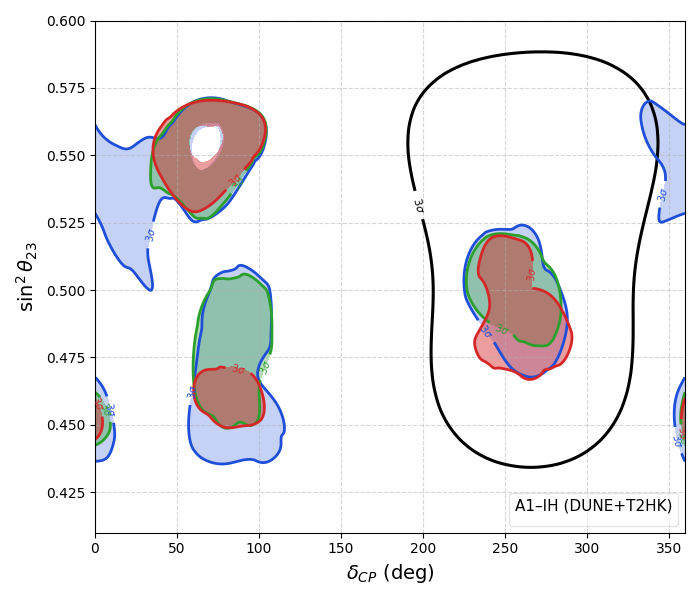}
    \label{fig:subC6}
  \end{subfigure}
  \begin{subfigure}[b]{0.45\textwidth}
    \centering
    \includegraphics[width=\textwidth]{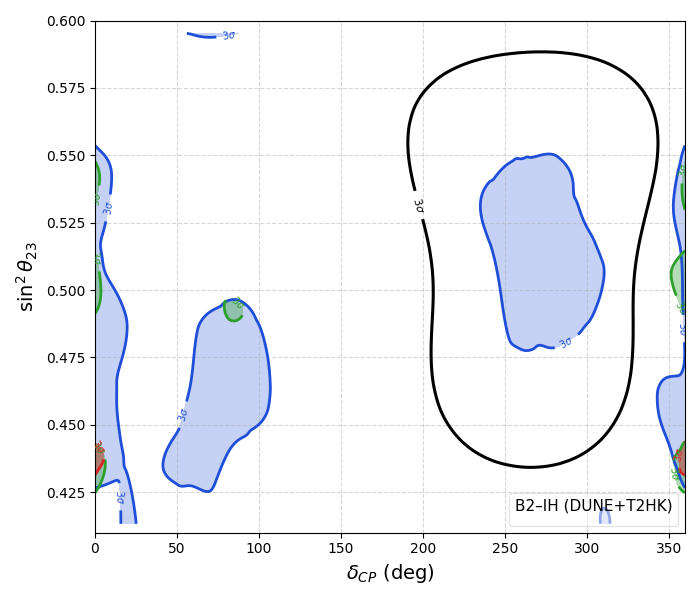}
    \label{fig:subD5}
  \end{subfigure}
  \\[-9ex]
  \begin{subfigure}[b]{0.45\textwidth}
    \centering
    \includegraphics[width=\textwidth]{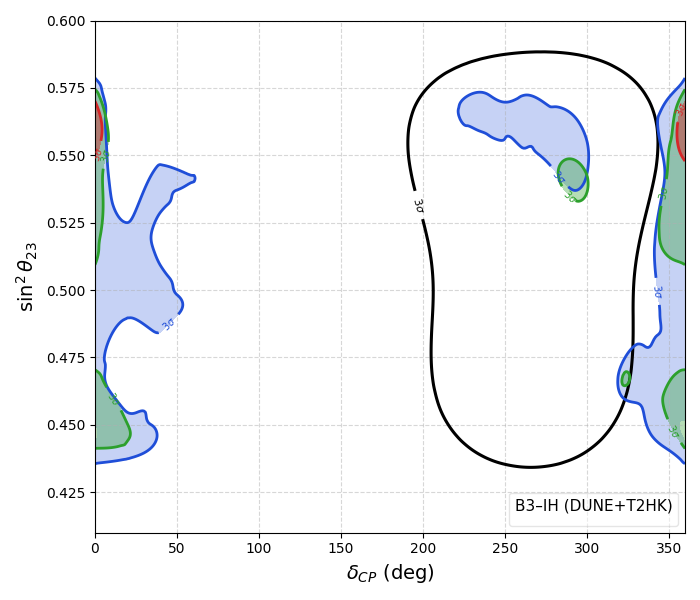}
    \label{fig:subC7}
  \end{subfigure}
  \begin{subfigure}[b]{0.45\textwidth}
    \centering
    \includegraphics[width=\textwidth]{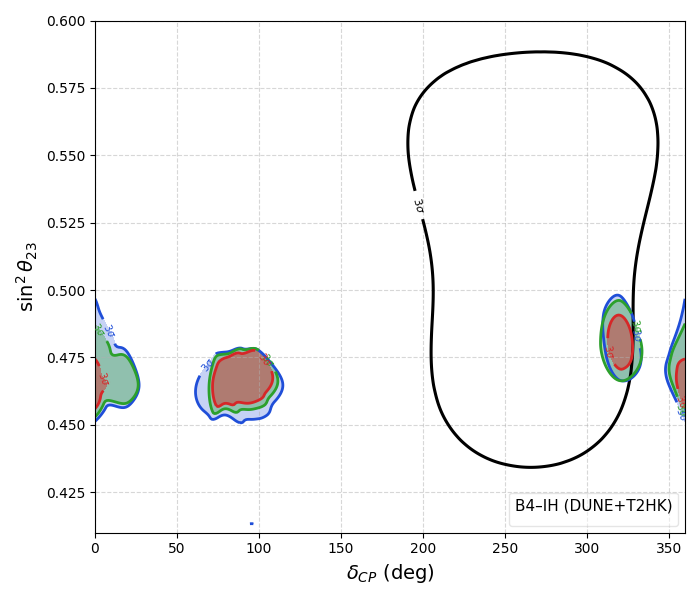}
    \label{fig:subD6}
  \end{subfigure}
  \\[-4ex]

  \caption{3$\sigma$ allowed regions in the $\theta_{23}$--$\delta_{\rm CP}$ plane for two-zero textures predicted by the modular $A_4$ left--right symmetric framework at DUNE+T2HK when assumed mass ordering is inverted. The colour code is same as Figure.~\ref{fig:NH}.}
  \label{fig:IH-comb}
\end{figure}

Compared to the fig.~\ref{fig:IH-dune}, the inclusion of T2HK leads to a
noticeable reduction of the allowed parameter space for most of the two-zero textures as seen from fig.~\ref{fig:IH-comb}. The allowed regions are now very much localized indication possible precision measurements. Once priors on $\theta_{12}$ and $\theta_{13}$ are imposed, allowed regions shrinks further. 

\textit{For some of the B-type textures, such as $B_{2}$ and $B_{3}$, the inclusion of priors on both
$\theta_{12}$ and $\theta_{13}$ significantly enhances their predictivity. In the combined
 analysis, the true parameter space collapses to highly restricted
regions, with most of the originally allowed $\theta_{23}$--$\delta_{\rm CP}$ combinations
becoming incompatible with the experimental capabilities. So with the precision inputs on $\theta_{12}$ and $\theta_{13}$, the combination of these experiments can prove $B_{1}$ and $B_{2}$ two-zero textures predicted by the modular $A_4$ left--right symmetric framework if inverted mass ordering is assumed to be the true mass order. }

\begin{figure}[!h]
    \centering
    \includegraphics[width=0.46\textwidth]{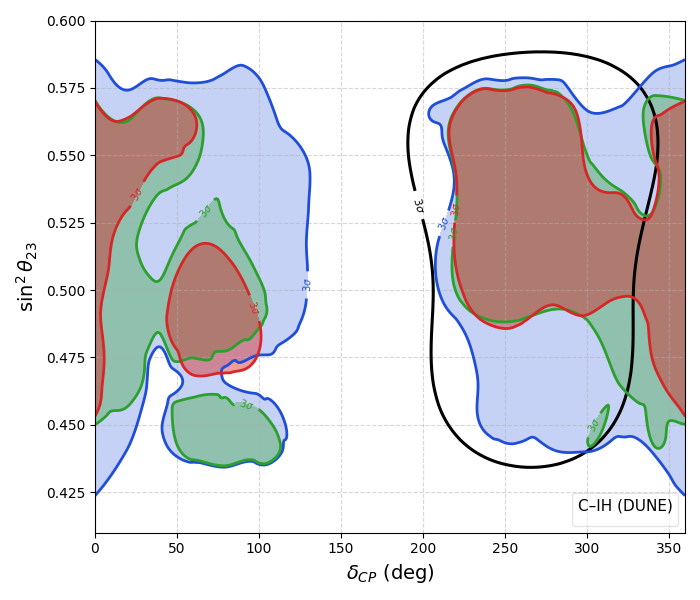}\hfill
    \includegraphics[width=0.46\textwidth]{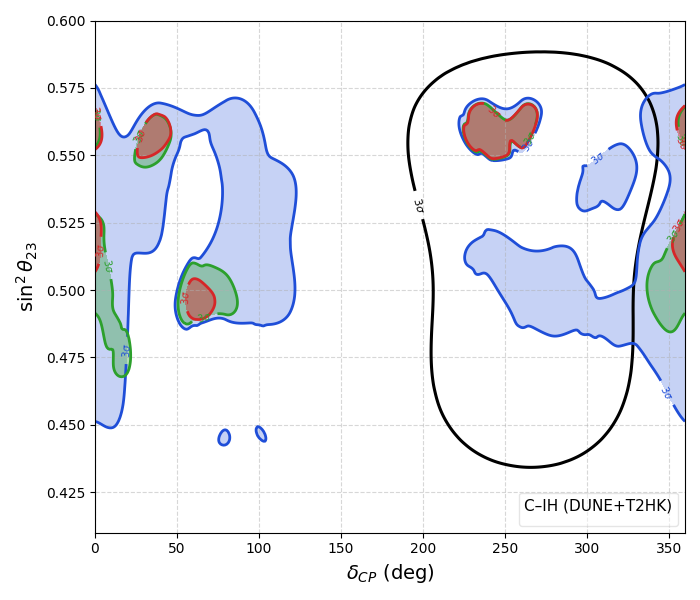}

    \caption{3$\sigma$ allowed regions in the $\theta_{23}$--$\delta_{\rm CP}$ plane for two-zero textures predicted by the modular $A_4$ left--right symmetric framework at DUNE (left) and DUNE+T2HK (right) when assumed mass ordering is inverted. The colour code is same as Figure.~\ref{fig:NH}.}
    \label{c-ih}
\end{figure}

In presence of the T2HK, DUNE with a runtime of 13 years can exclude most of the $\theta_{23}$--$\delta_{\rm CP}$ parameter space in the IH mode for all the other textures including the C type (\ref{c-ih}) texture.

\subsection{Discussion on Effective Majorana Mass vs. Lightest Neutrino Mass}
\begin{figure}[H]
    \centering
    \includegraphics[width=0.45\textwidth, height=5cm]{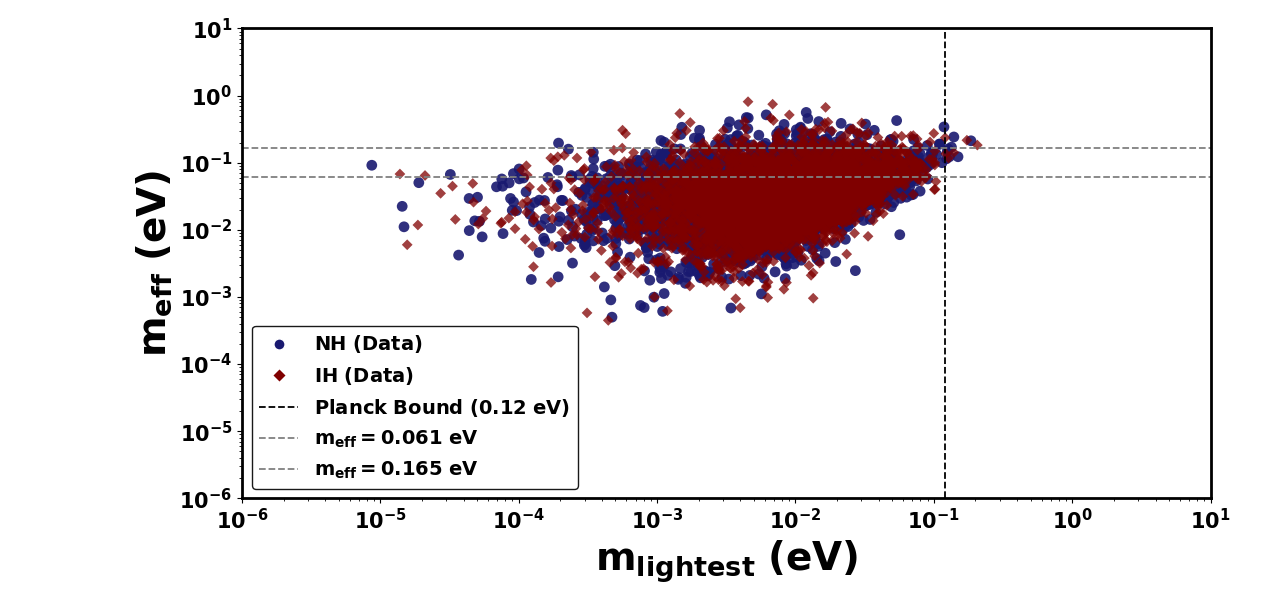}
    \includegraphics[width=0.45\textwidth, height=5cm]{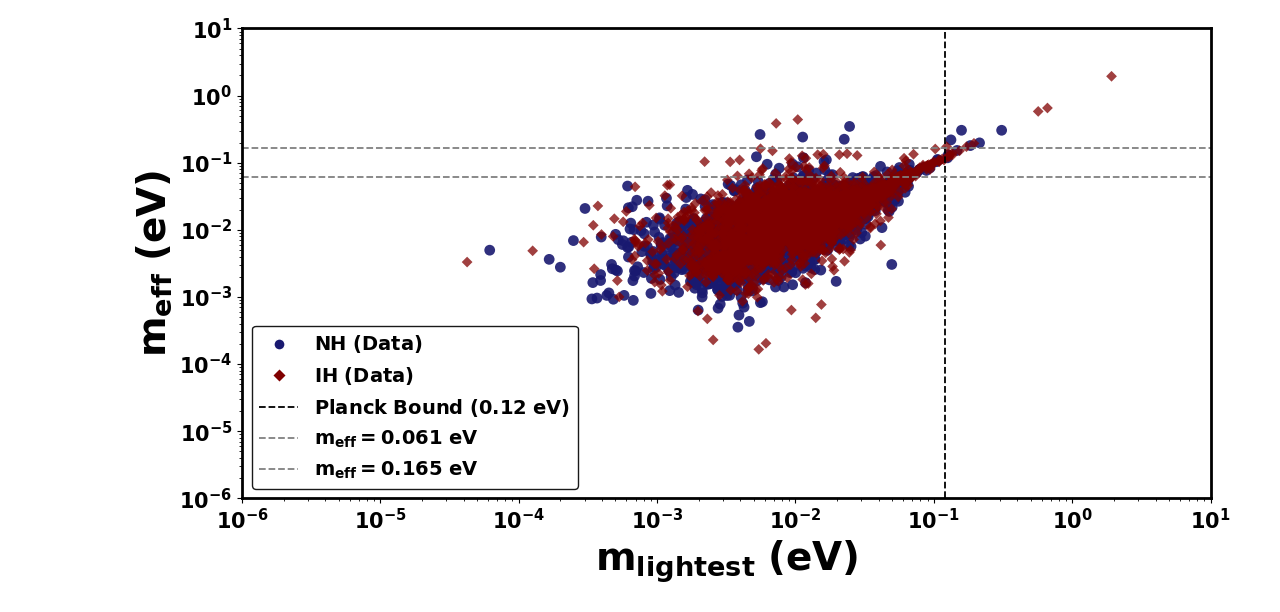}
    
    \caption{Variation of the effective Majorana mass \(|m^{\nu}_{\text{eff}}|\) with respect to the lightest neutrino mass for both NH and IH. 
    The left panel corresponds to Texture $B_{4}$ and the right panel to Texture $C$. 
    The plot includes the experimentally allowed regions relevant for neutrinoless double beta decay.}
    \label{W3F11}
\end{figure}

  We have calculated the effective Majorana mass, $m_{\text{eff}}$, for the standard contribution. Since the $(1,1)$ element of the neutrino mass matrix is zero in the case of textures $A_{1}$ and $A_{2}$, the effective Majorana mass $m_{\text{eff}}$ is zero for these two textures. We have calculated $m_{\text{eff}}$ from the model for the remaining five two-zero textures for both  NH and IH. Figure \ref{W3F11} shows the variation of effective mass with respect to the lightest neutrino mass for texture $B_{4}$ and C. In all the figures, the vertical line represents the Planck bound on the $\sum m_{\nu}$. The two horizontal lines in each figure correspond to experimental bounds, which arise due to uncertainties in the calculation of the nuclear matrix elements. The effective Majorana mass should lie below these experimental limits.
\subsection{Branching Ratio Analysis: Results }
We have calculated the branching ratio for three lepton flavor violating (LFV) processes, namely $\tau \rightarrow e\gamma$, $\tau \rightarrow \mu\gamma$, and $\mu \rightarrow e\gamma$. The experimental upper bounds on the branching ratios are given in Table~\ref{W3T10}. We have plotted the branching ratios against the lightest neutrino mass for both NH and IH. All the two-zero textures are capable of yielding branching ratios within the experimental limits. We found that for all seven two-zero textures, the calculated branching ratio for the decay $\mu \rightarrow e\gamma$ is well below the experimental bound. Each subplot corresponds to one of the LFV processes and includes the respective experimental upper bounds for comparison.
\begin{table}[H]
\centering
\begin{tabular}{|c|c|c|}
\hline
Branching ratio for LFV processes & Experimental bounds \\ \hline
$ Br(\tau \rightarrow e\gamma)$ & $< 1.5 \times 10^{-8}$ \cite{BaBar:2009hkt}  \\ \hline
$Br(\tau \rightarrow \mu\gamma)$ & $< 1.5 \times 10^{-8}$ \cite{BaBar:2009hkt} \\ \hline
$Br(\mu \rightarrow e\gamma)$ & $<4.2 \times 10^{-13}$ \cite{MEG:2016leq} \\ \hline
\end{tabular}
\caption{Experimental upper bound on LFV process}
\label{W3T10}
\end{table}
\begin{figure}[H]
    \centering
  
    \includegraphics[width=0.32\textwidth, height=4cm]{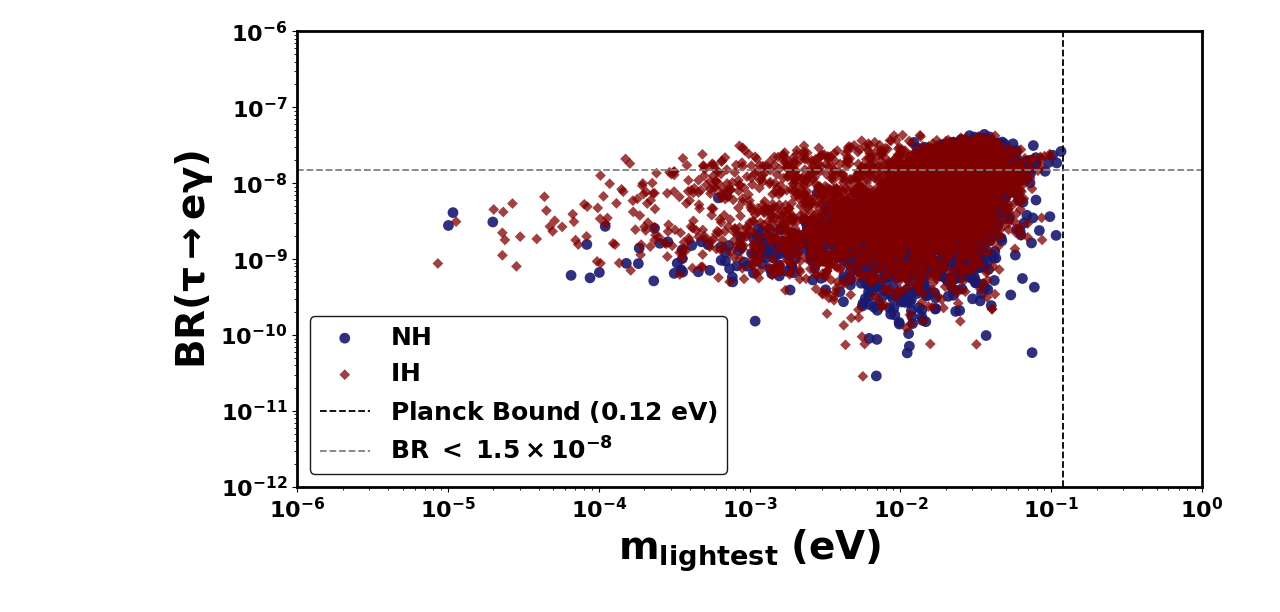}
    \includegraphics[width=0.32\textwidth, height=4cm]{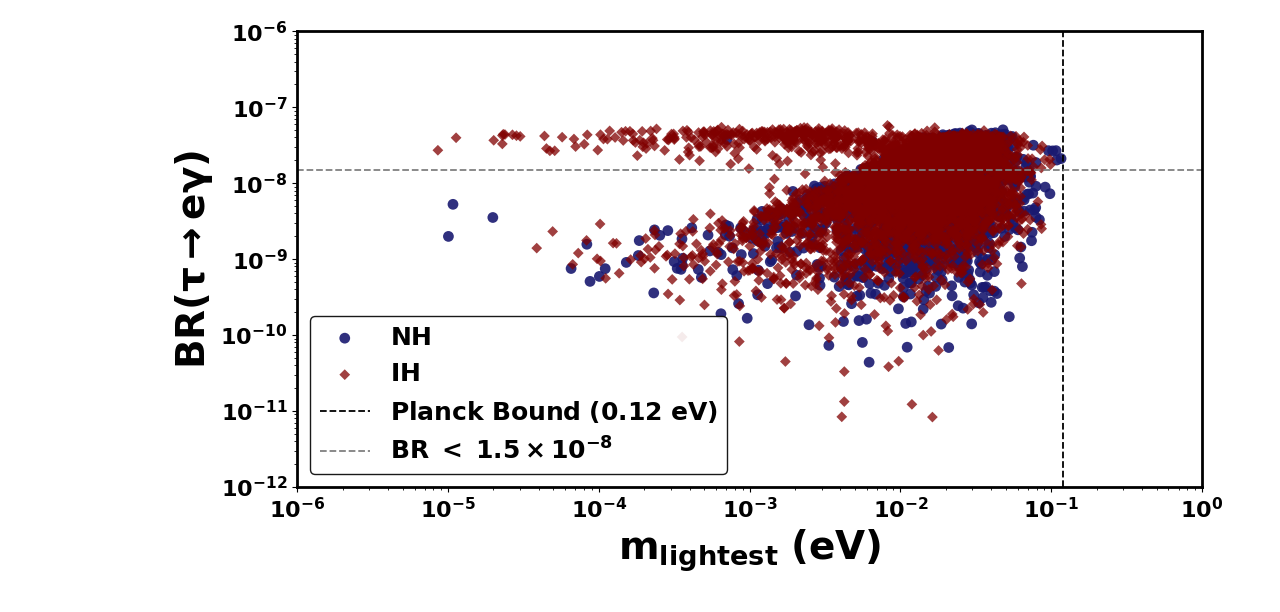}
    \includegraphics[width=0.32\textwidth, height=4cm]{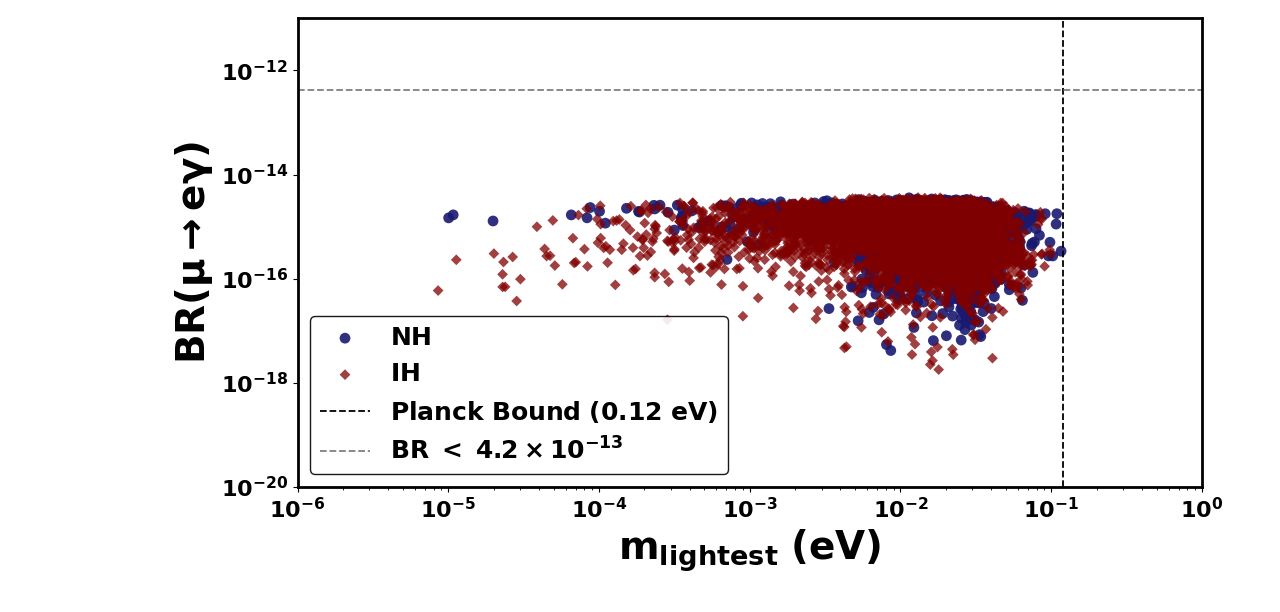}
    
    \includegraphics[width=0.32\textwidth, height=4cm]{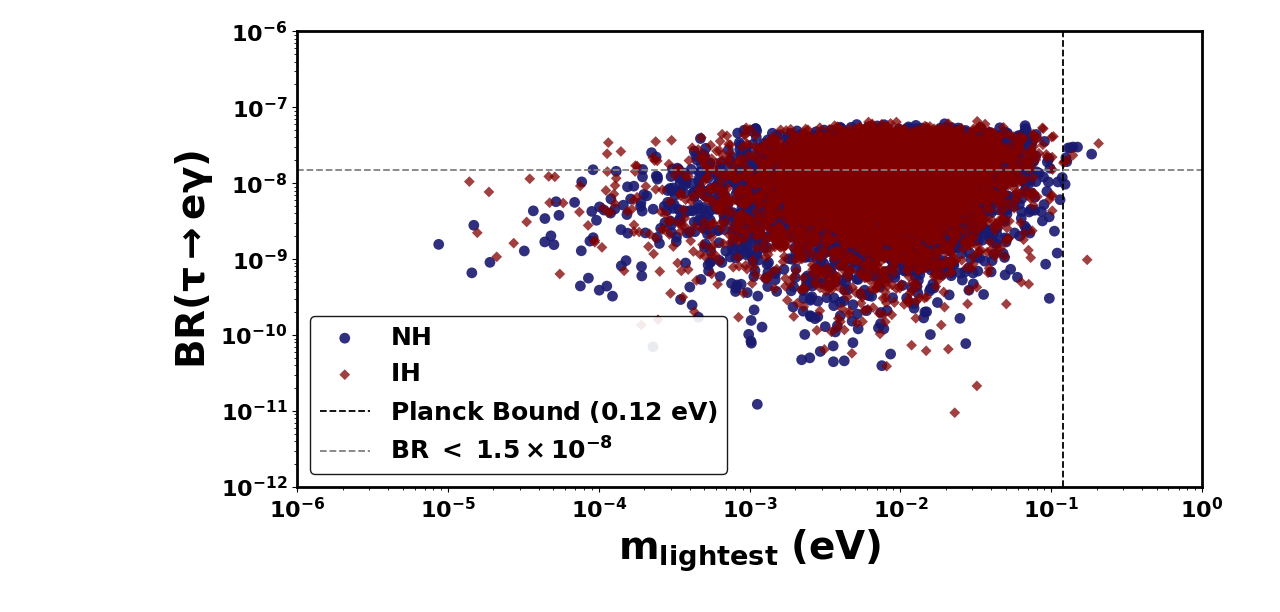}
    \includegraphics[width=0.32\textwidth, height=4cm]{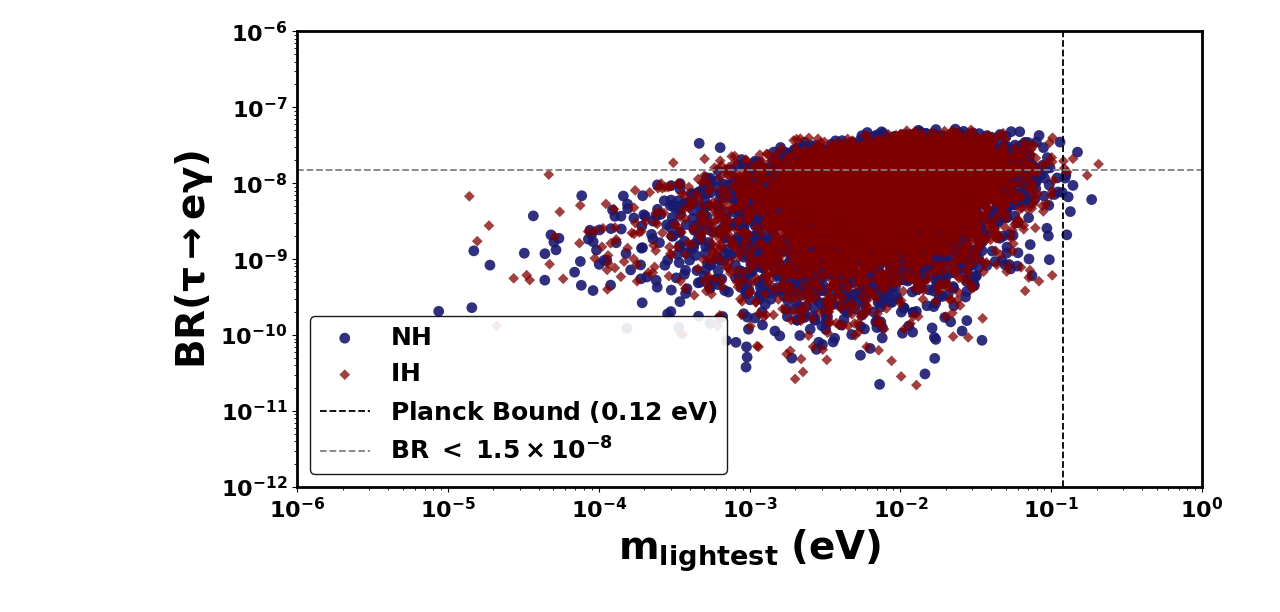}
    \includegraphics[width=0.32\textwidth, height=4cm]{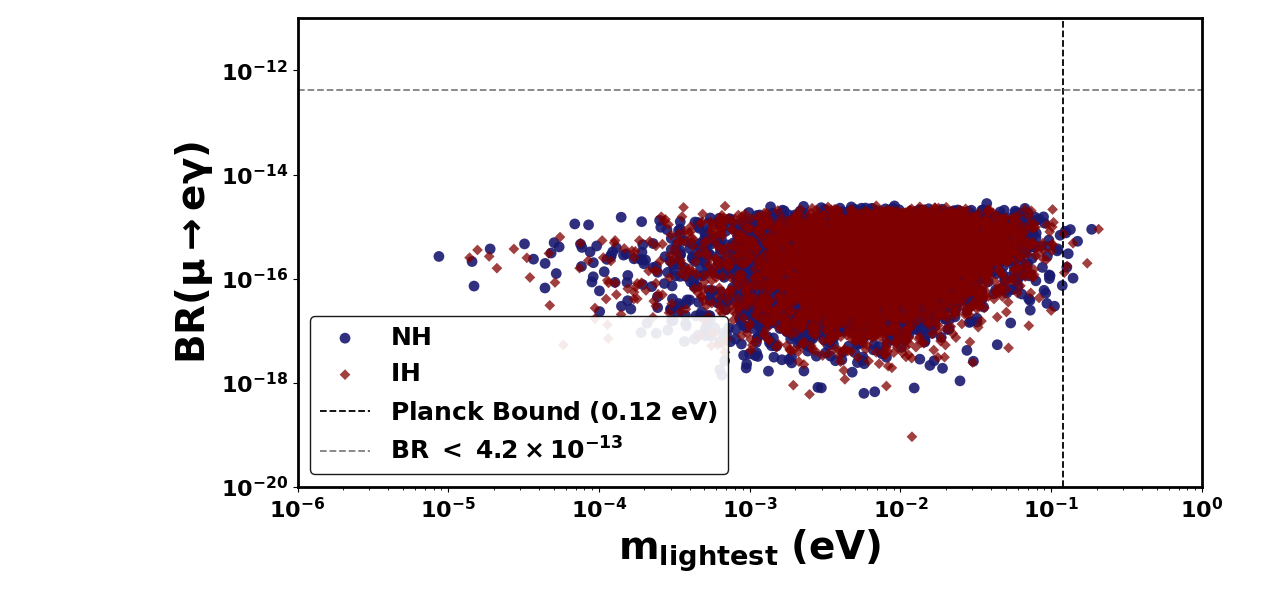}
    
    \includegraphics[width=0.32\textwidth, height=4cm]{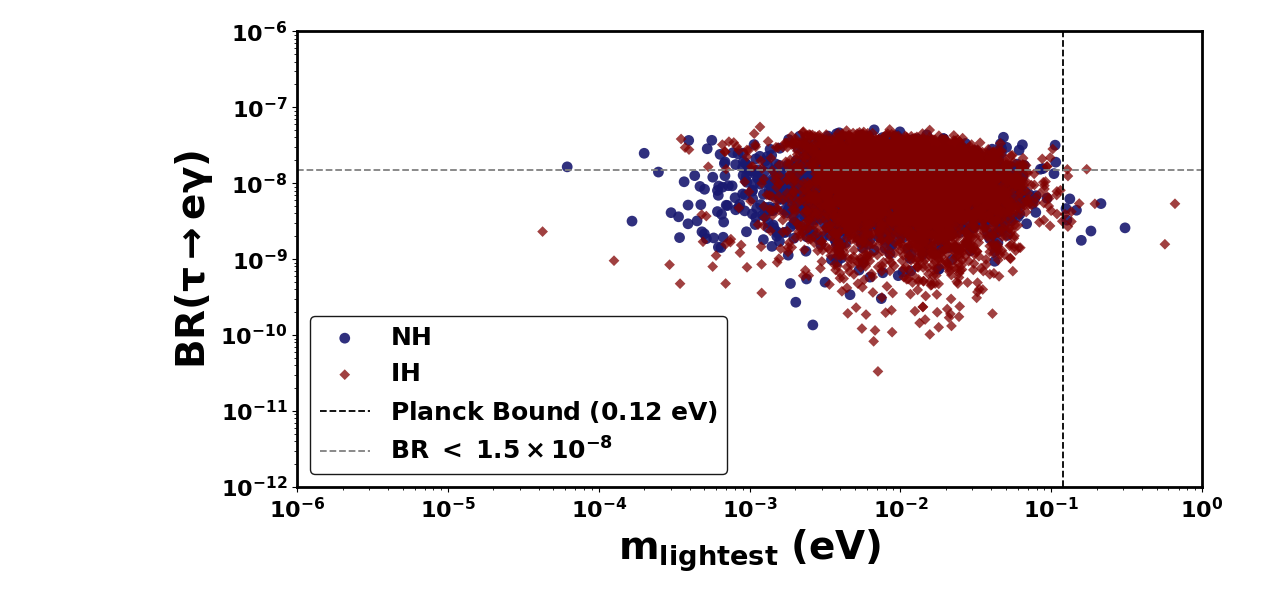}
    \includegraphics[width=0.32\textwidth, height=4cm]{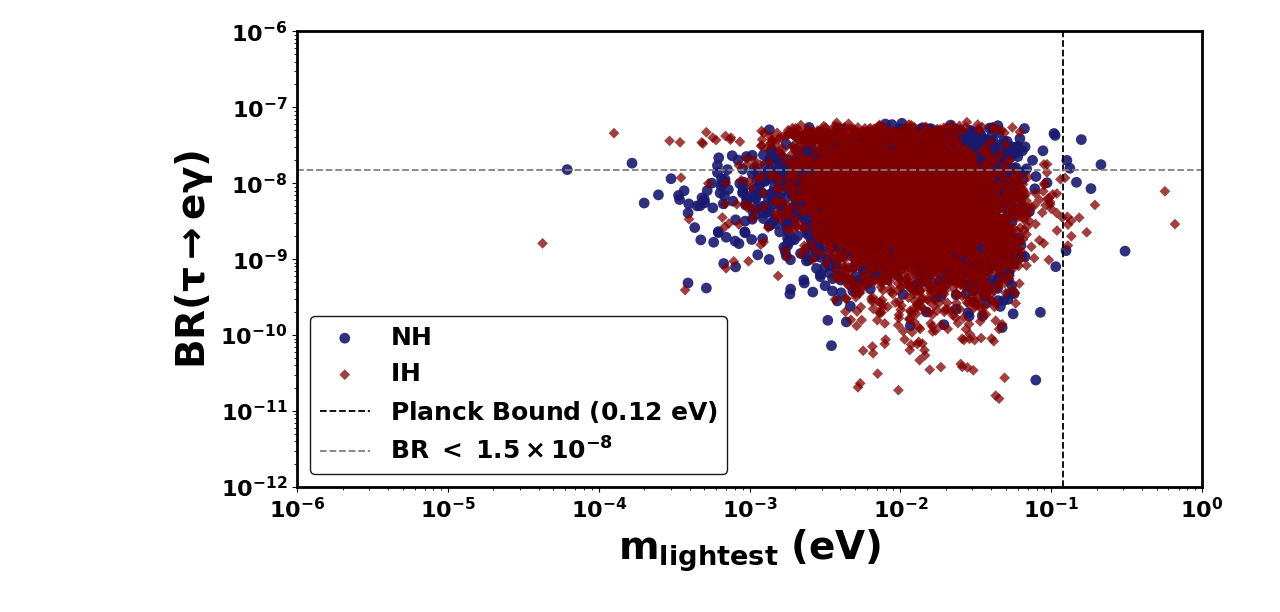}
    \includegraphics[width=0.32\textwidth, height=4cm]{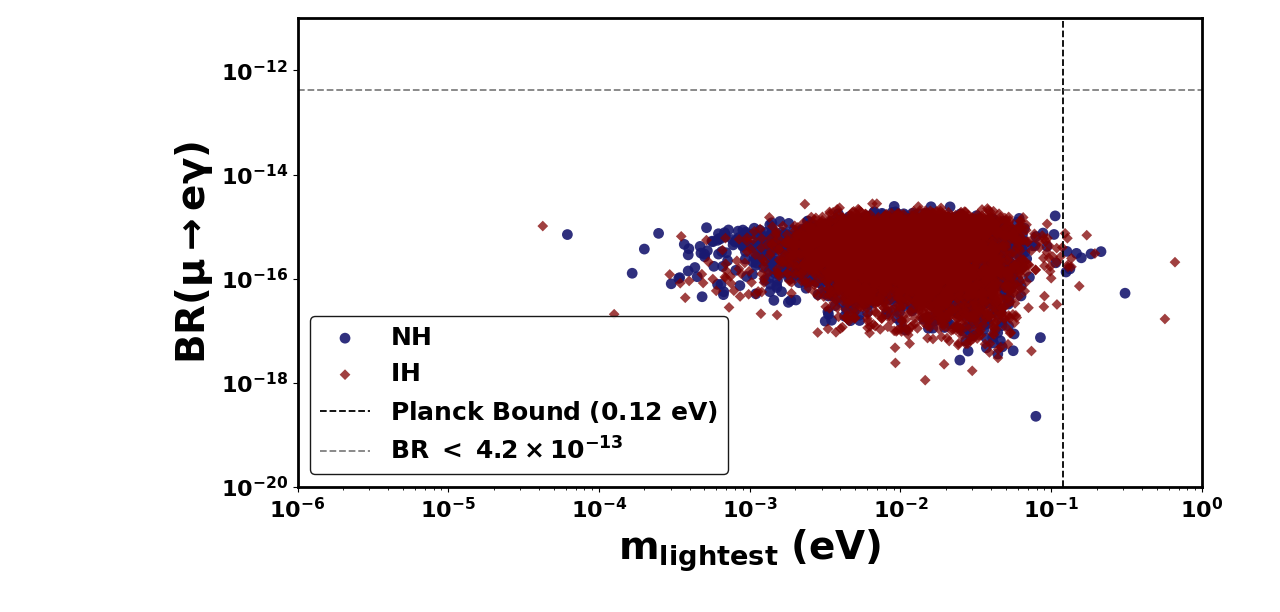}
    
    \caption{Variation of the branching ratios for the lepton flavor violating processes with respect to the lightest neutrino mass for both NH and IH. 
    The first row corresponds to Texture $A_{1}$, the second row to Texture $B_{4}$, and the third row to Texture C.}
    \label{fig:BR_textures}
\end{figure}
\begin{table}[H]
\centering
\scriptsize  
\setlength{\tabcolsep}{12pt} 
\renewcommand{\arraystretch}{1.3} 
\begin{tabular}{|c|c|c|c|c|c|c|c|c|}
\hline
\textbf{Observable} & \textbf{Hierarchy} & \textbf{A$_1$} & \textbf{A$_2$} & \textbf{B$_1$} & \textbf{B$_2$} & \textbf{B$_3$} & \textbf{B$_4$} & \textbf{C} \\
\hline
$m^{\nu}_{\text{eff}}$ (eV) & NH & \checkmark & \checkmark & \checkmark & \checkmark & \checkmark & \checkmark & \checkmark \\
                            & IH & \checkmark & \checkmark & \checkmark & \checkmark & \checkmark & \checkmark & \checkmark \\
\hline
BR($\mu \rightarrow e\gamma$) & NH & \checkmark & \checkmark & \checkmark & \checkmark & \checkmark & \checkmark & \checkmark \\
                              & IH & \checkmark & \checkmark & \checkmark & \checkmark & \checkmark & \checkmark & \checkmark \\
\hline
BR($\tau \rightarrow e\gamma$) & NH & \checkmark & \checkmark & \checkmark & \checkmark & \checkmark & \checkmark & \checkmark \\
                               & IH & \checkmark & \checkmark & \checkmark & \checkmark & \checkmark & \checkmark & \checkmark \\
\hline
BR($\tau \rightarrow \mu\gamma$) & NH & \checkmark & \checkmark & \checkmark & \checkmark & \checkmark & \checkmark & \checkmark \\
                                 & IH & \checkmark & \checkmark & \checkmark & \checkmark & \checkmark & \checkmark & \checkmark \\
\hline
\end{tabular}
\caption{Summary of results for different two-zero textures under NH and IH}
\label{W3T11}
\end{table}
\section{Conclusion}\label{s6}
In this work, we have realized all seven experimentally compatible two-zero textures of the neutrino mass matrix using the $\Gamma_{3}$ modular group. The light neutrino mass matrix is generated via an extended Type II seesaw mechanism, and to realize different two-zero textures, the charged leptons are assigned singlet representations of the $A_{4}$ group. Modular weight and charge assignments are given to the supermultiplets in such a way that specific elements of the neutrino mass matrix do not appear in the superpotential and can thus be treated as zero.\\

All seven two-zero textures are found to be compatible with current oscillation data for
both normal and inverted mass orderings, with no strong preference for either hierarchy. The predicted values of the effective Majorana mass for textures $B_{1}, B_{2}, B_{3}, B_{4}$, and $C$ lie approximately in the range $10^{-1}$ to $10^{-4}$ eV. From the plots of $m_{\text{eff}}$ versus the lightest neutrino mass, we observed that when the lightest neutrino mass varies from $10^{-1}$ eV to $10^{-5}$ eV, the effective mass remains within the experimentally allowed region, indicating consistency with current
neutrinoless double beta decay constraints.\\
Importantly, the highly restricted correlations among the neutrino oscillation parameters as predicted by the considered framework, have been tested at long-baseline neutrino experiments.
As demonstrated by our detailed analysis at DUNE and DUNE+T2HK
setup, the inclusion of precision external constraints on $\theta_{12}$ and $\theta_{13}$
substantially enhances the ability of future experiments to discriminate among the
different two-zero textures and, in several cases, to exclude large fractions of the
globally allowed parameter space. When priors on both $\theta_{12}$ and $\theta_{13}$ are imposed with the combined set-up, the allowed regions
shrink significantly and collapse into compact, well-localized islands in the
$\theta_{23}$--$\delta_{\rm CP}$ plane, resulting in a substantial enhancement in the
predictability of the texture predictions. or inverted mass ordering, the DUNE--T2HK synergy leads to strong predictivity for the $B_{2}$ and $B_{4}$ textures, with the surviving regions confined to tiny islands around the CP-conserving solutions in the lower and higher octant of $\theta_{23}$, respectively.

Overall, this study demonstrate that the synergy between DUNE and T2HK, when accounted with the high-precision external constraints on $\theta_{12}$ and $\theta_{13}$, can significantly
enhances the capability of long-baseline neutrino experiments to test, constrain, and
potentially falsify two-zero neutrino mass textures.

\acknowledgments
D.D. acknowledges support from the Focus Area Science Technology Summer Fellowship Programme of the three National Science Academies of India for a research visit to the Physical Research Laboratory, Ahmedabad.

\revappendix
\section{Results of Effective Majorana Mass and Its Dependence on Lightest Neutrino Mass}\label{W3A4}
\begin{figure}[H]
    \centering
    \includegraphics[width=0.32\textwidth, height=4cm]{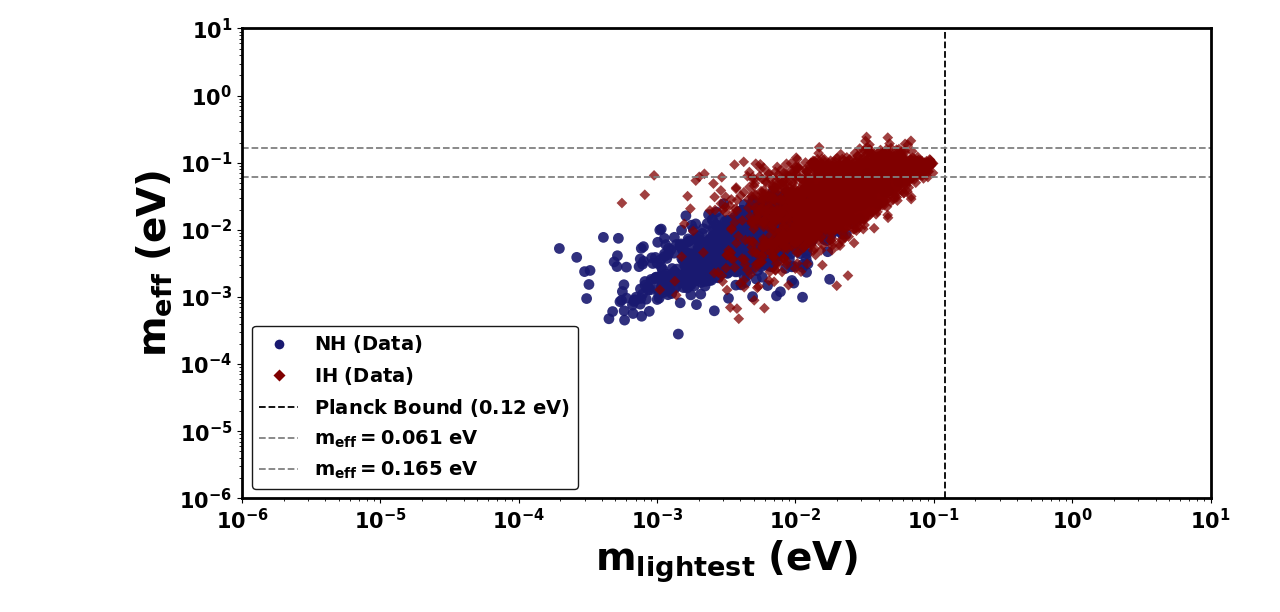}
    \includegraphics[width=0.32\textwidth, height=4cm]{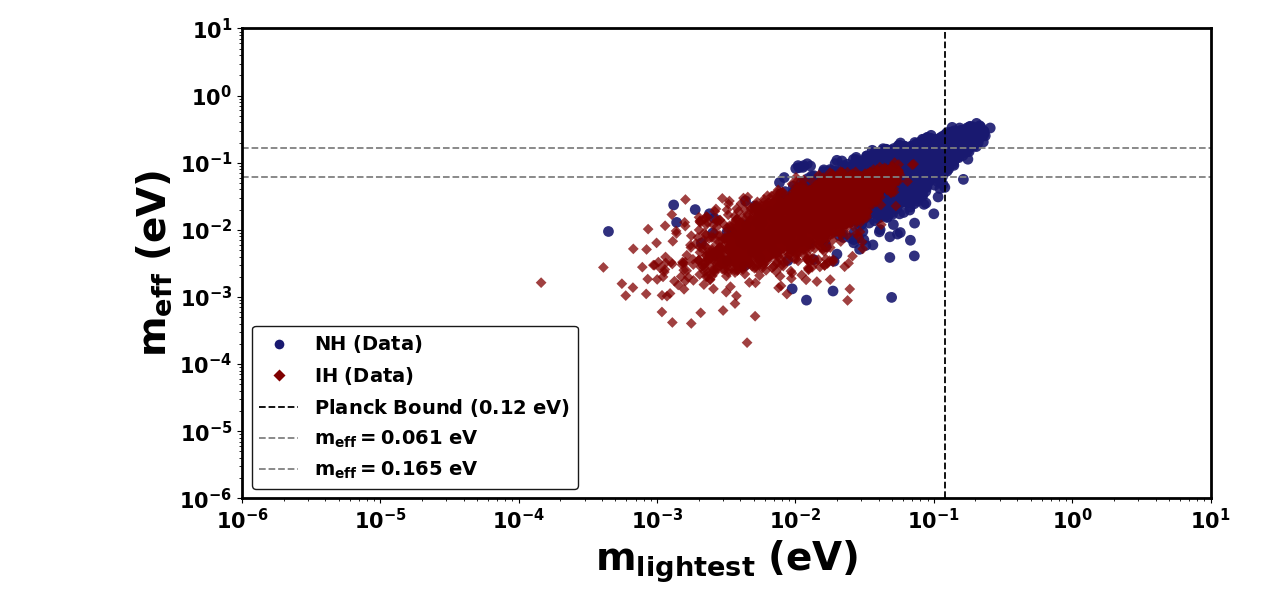}
    \includegraphics[width=0.32\textwidth, height=4cm]{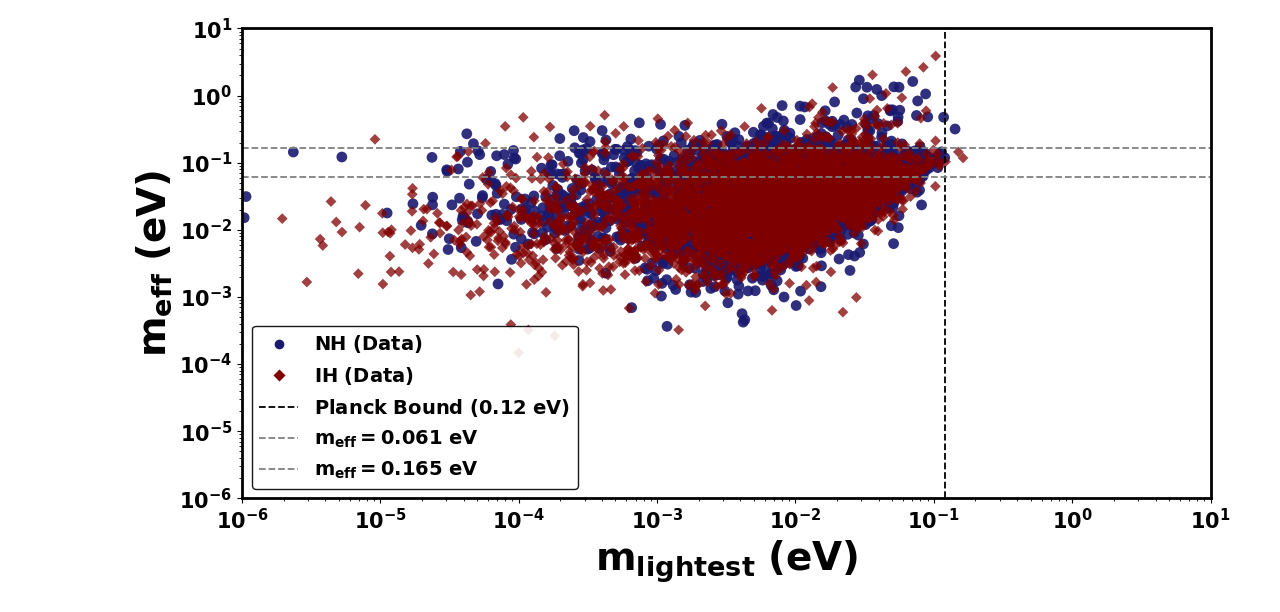}
    \caption{Variation of the effective Majorana mass \(|m^{\nu}_{eff}|\) with respect to the lightest neutrino mass for both NH and IH. The row corresponds to cases $B_{1}$, $B_{2}$, and $B_{3}$ (from left to right).}
    \label{W3F21}
\end{figure}

\section{Calculation of Branching Ratio and Its Correlation with Lightest Neutrino Mass}\label{W3A5}
\begin{figure}[H]
    \centering
    \includegraphics[width=0.32\textwidth, height=4cm]{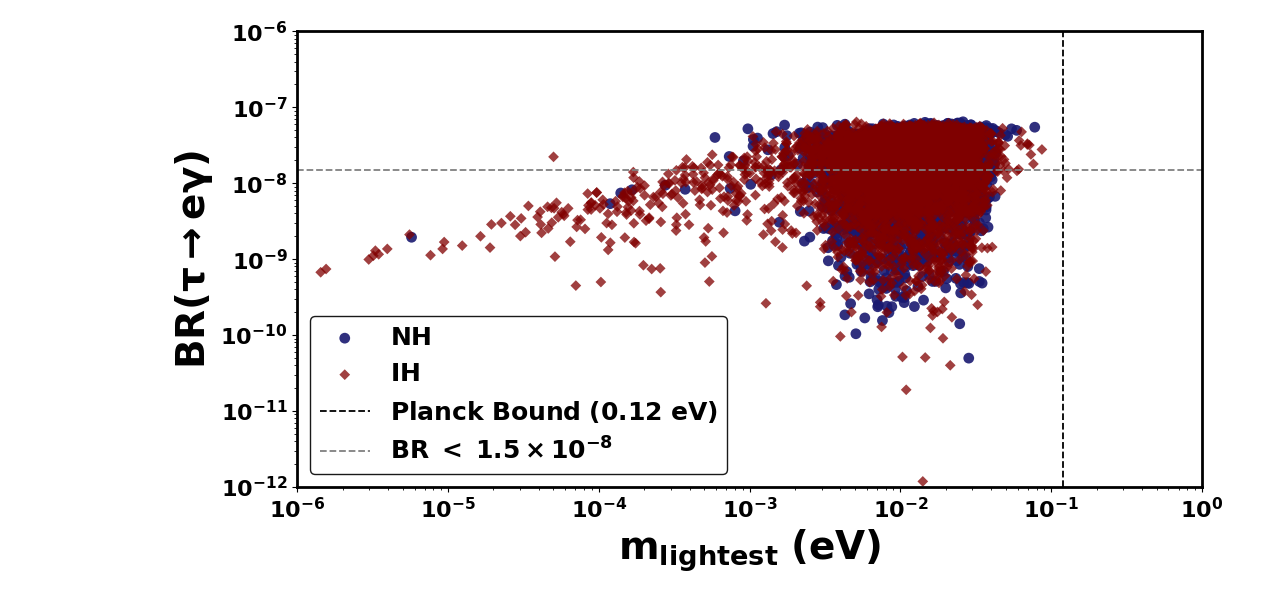}
    \includegraphics[width=0.32\textwidth, height=4cm]{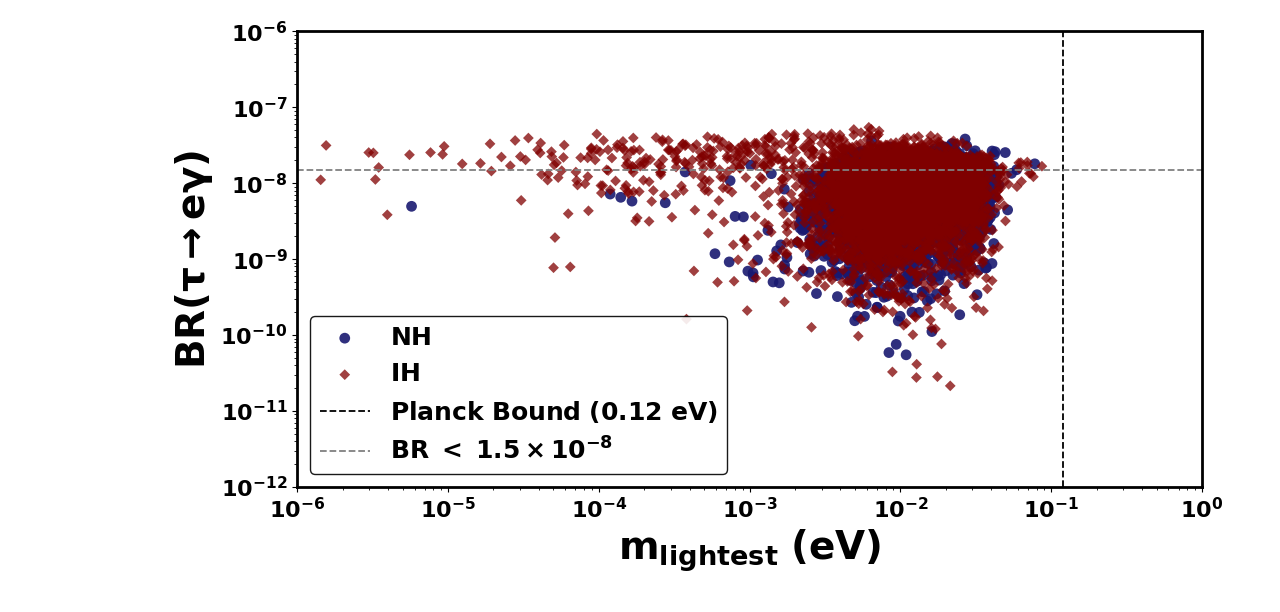}
    \includegraphics[width=0.32\textwidth, height=4cm]{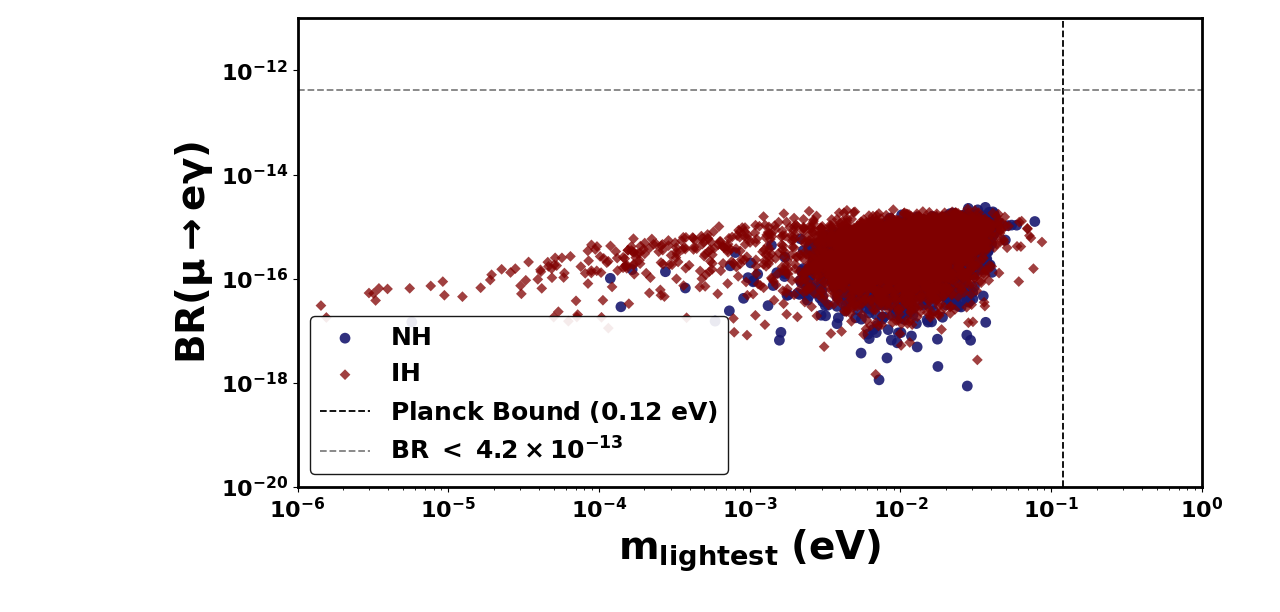}

    \includegraphics[width=0.32\textwidth, height=4cm]{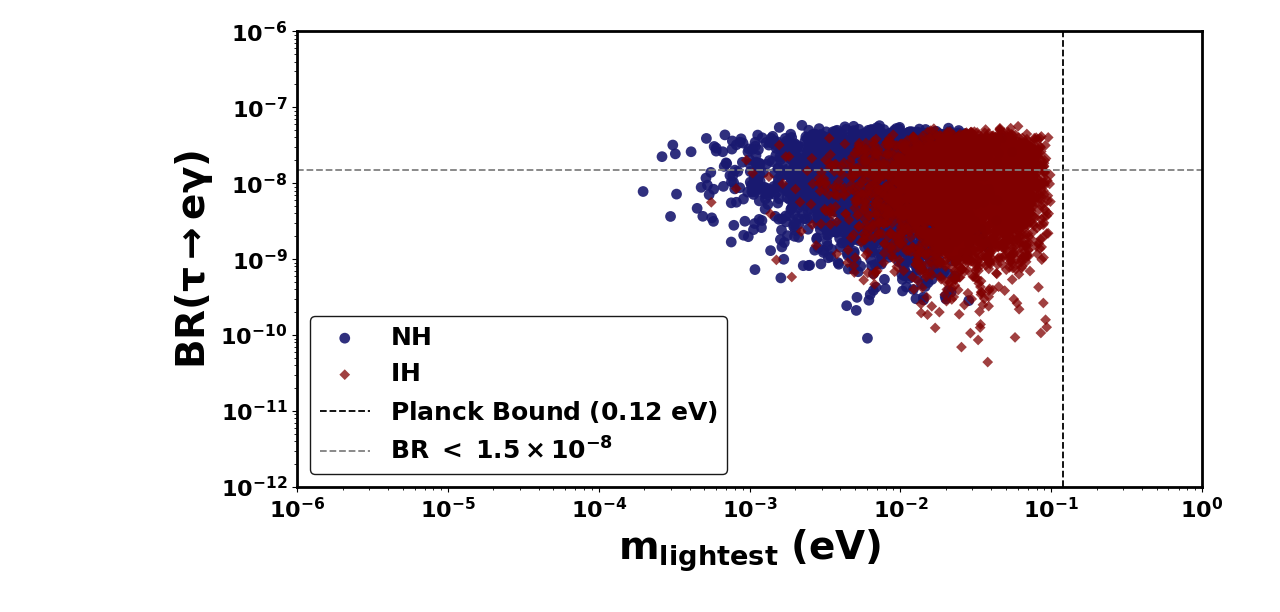}
    \includegraphics[width=0.32\textwidth, height=4cm]{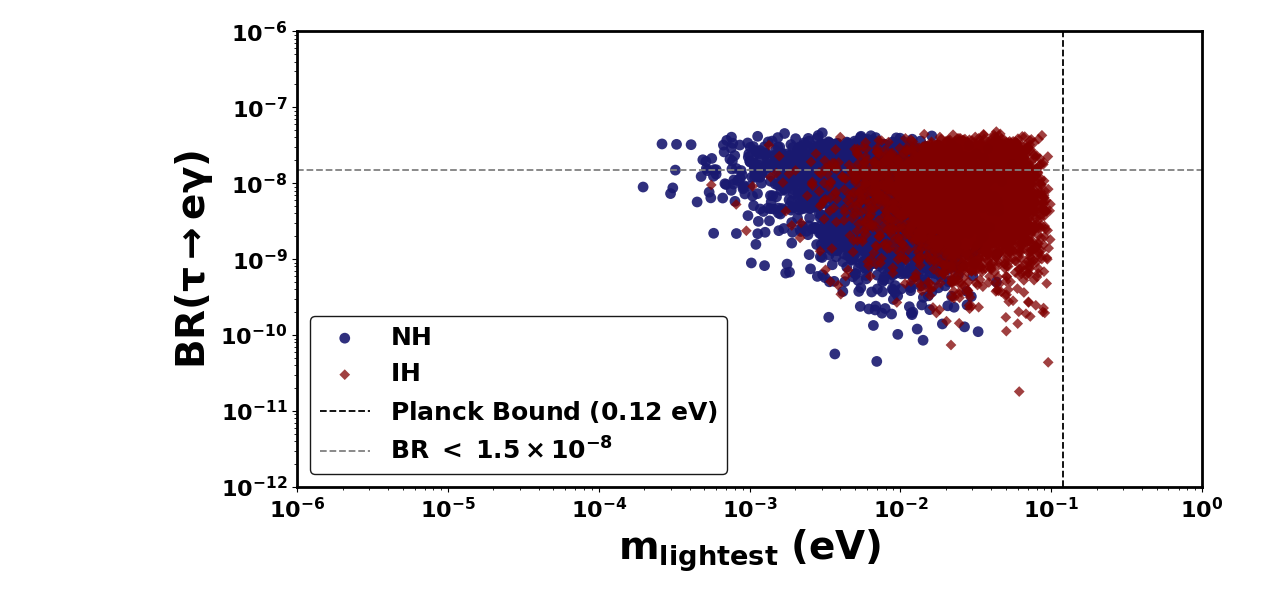}
    \includegraphics[width=0.32\textwidth, height=4cm]{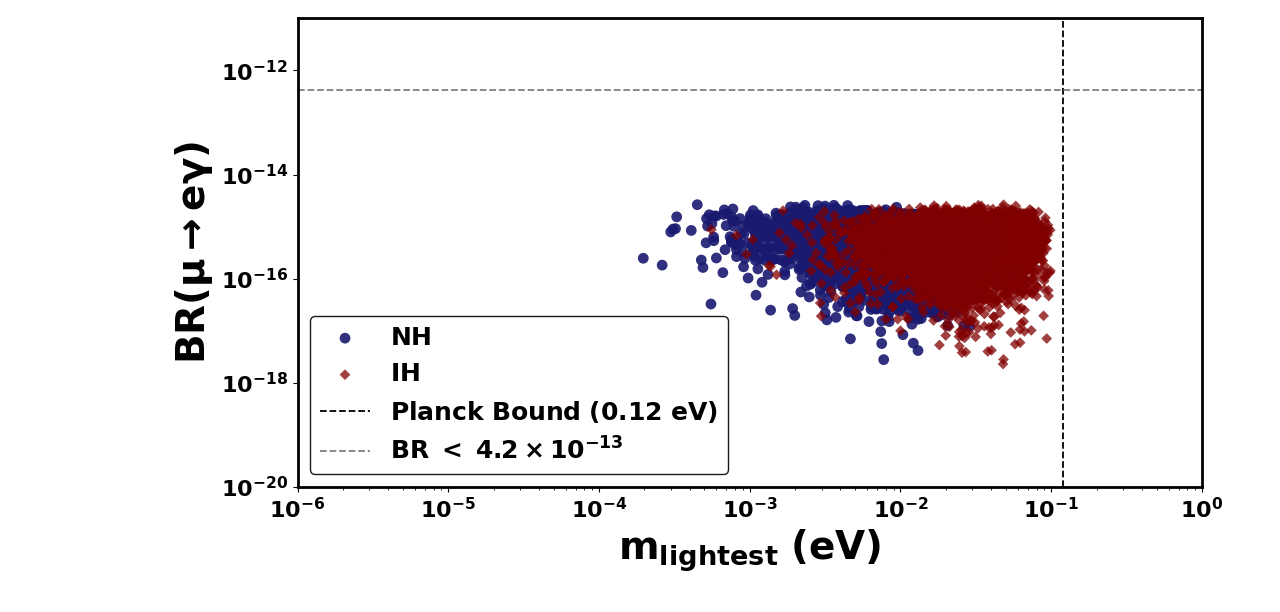}

    \includegraphics[width=0.32\textwidth, height=4cm]{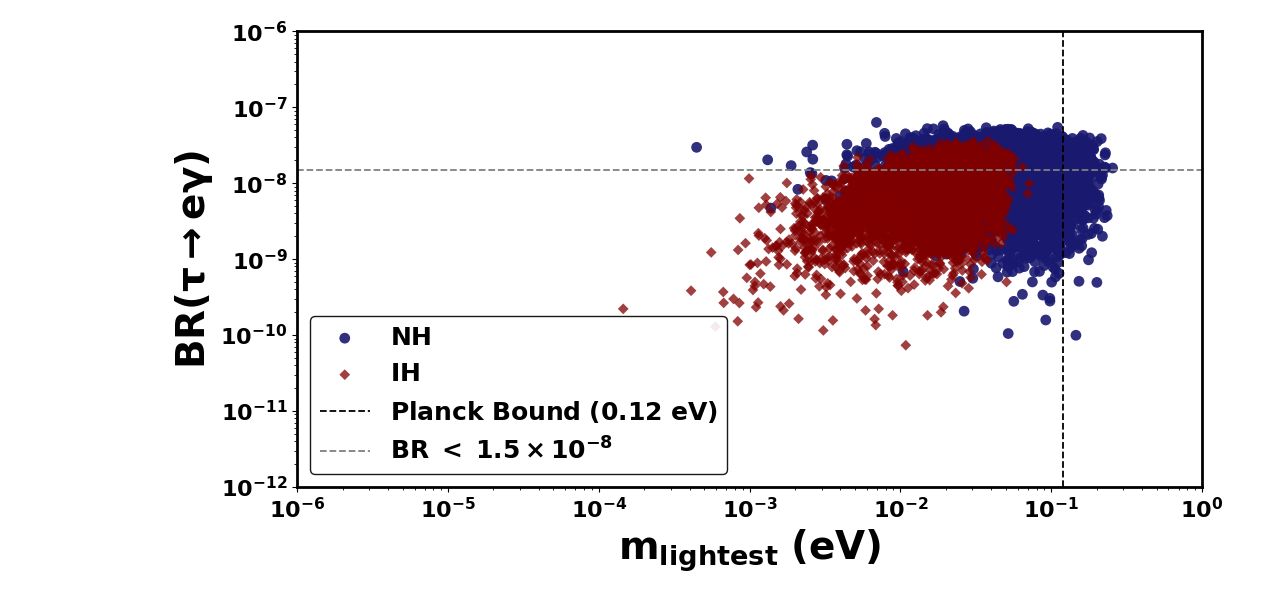}
    \includegraphics[width=0.32\textwidth, height=4cm]{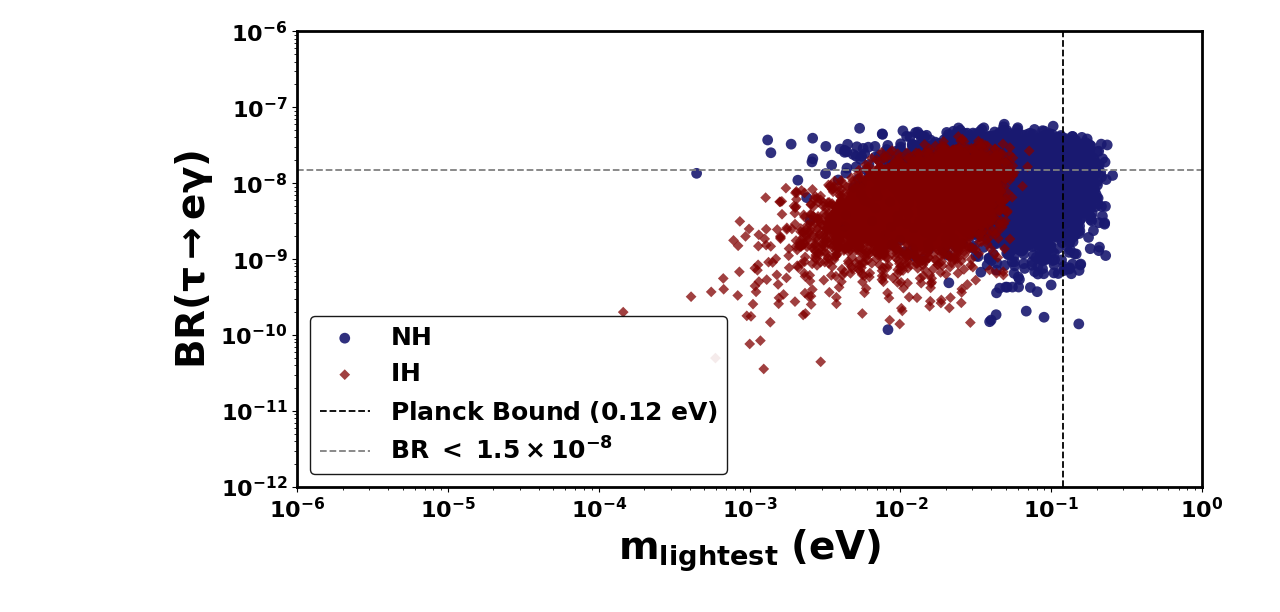}
    \includegraphics[width=0.32\textwidth, height=4cm]{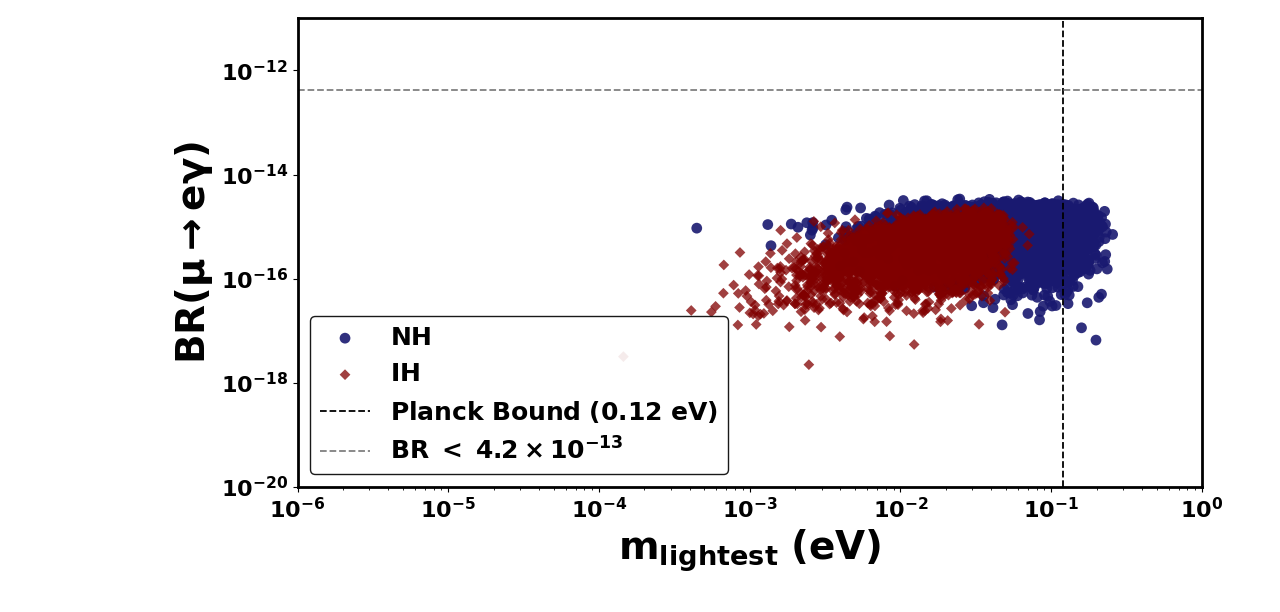}

    \includegraphics[width=0.32\textwidth, height=4cm]{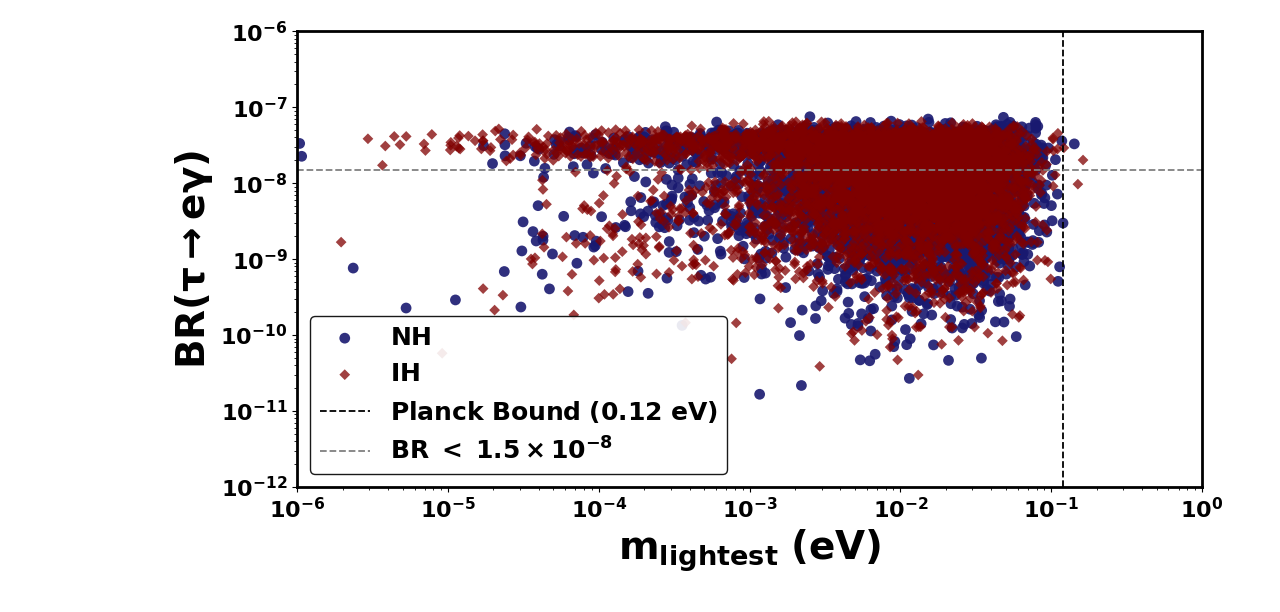}
    \includegraphics[width=0.32\textwidth, height=4cm]{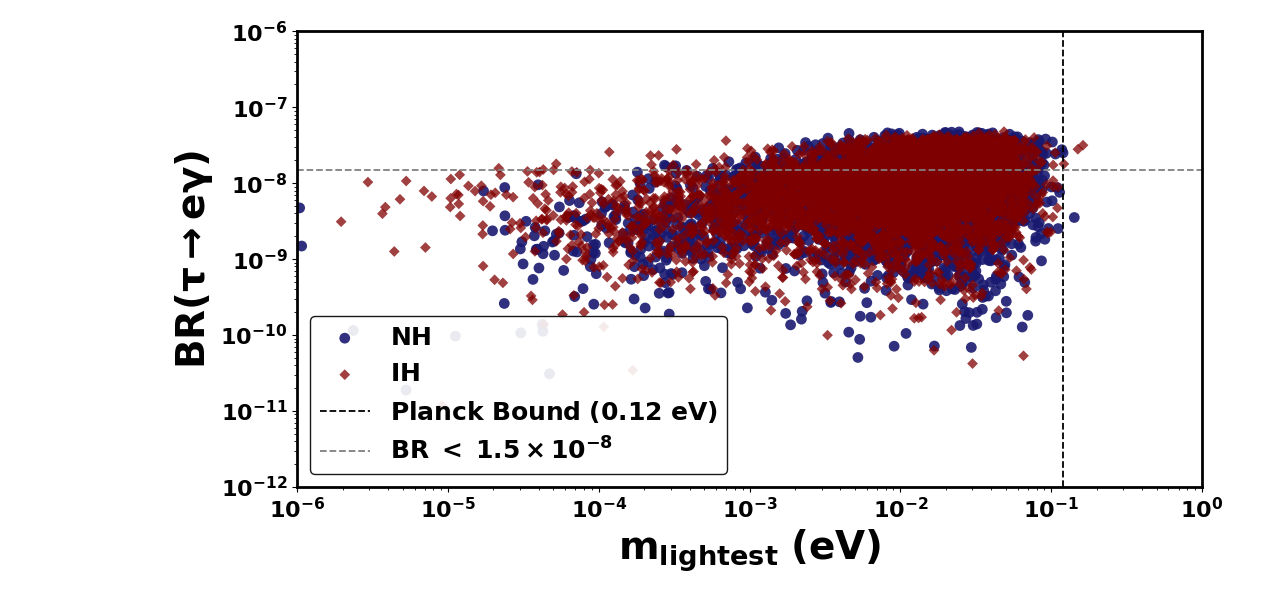}
    \includegraphics[width=0.32\textwidth, height=4cm]{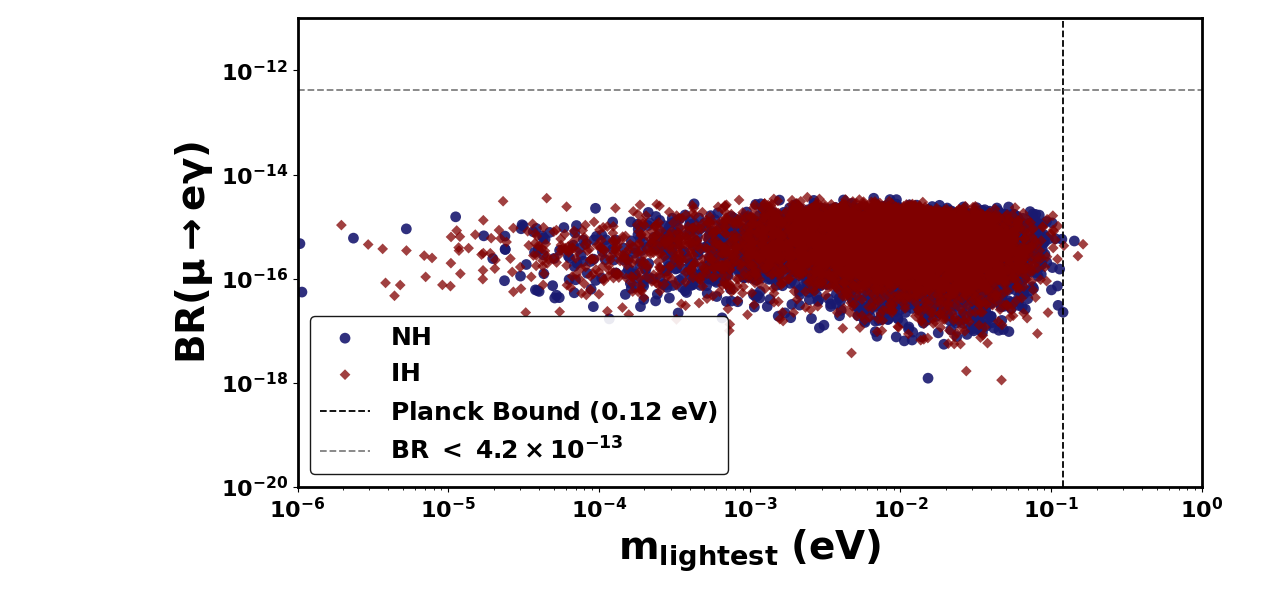}

    \caption{Variation of the branching ratios for the lepton flavor violating processes with respect to the lightest neutrino mass for both NH and IH. 
    The first row corresponds to Texture $A_{2}$, the second row to Texture $B_{1}$, the third row to Texture $B_{2}$, 
    and the fourth row to Texture $B_{3}$ (from left to right in each row).}
    \label{W3F20}
\end{figure}
\bibliographystyle{unsrt}
\bibliography{cite}
\end{document}